\documentclass[12pt,a4paper,english]{article}
\usepackage{titling}
\usepackage[affil-it]{authblk}
\usepackage[dvipsnames]{xcolor}
\usepackage{longtable}
\usepackage{lscape}
\usepackage{graphicx}
\usepackage{epsfig}
\usepackage{dsfont}
\usepackage{amsfonts}
\usepackage{amsmath}
\usepackage{amsthm}
\usepackage{amssymb}
\usepackage{bbold}
\usepackage[english]{babel}
\usepackage[utf8]{inputenc}
\usepackage{calc}
\usepackage{cancel}
\usepackage{float}
\usepackage{slashed}
\usepackage{mathrsfs}
\usepackage{pdfpages}
\usepackage{tabu}
\usepackage{simplewick}
\usepackage[hidelinks]{hyperref}
\usepackage{multirow}
\usepackage{hhline}
\usepackage{float}
\usepackage{bm}
\usepackage{braket}
\usepackage{comment}
\usepackage{cite}
\usepackage[left=0.8in,right=0.8in,top=1.0in,bottom=1.2in]{geometry}
\usepackage{datetime}
\usepackage{booktabs}
\usepackage{hyperref}
\date{}

\newcommand{\be}{\begin{equation}}
\newcommand{\ee}{\end{equation}}
\newcommand{\ba}{\begin{array}}
\newcommand{\ea}{\end{array}}
\newcommand{\beq}{\begin{equation}}
\newcommand{\eeq}{\end{equation}}
\newcommand{\beqa}{\begin{array}}
\newcommand{\eeqa}{\end{array}}
\newcommand{\bim}{\begin{itemize}}
\newcommand{\eim}{\end{itemize}}
\newcommand{\bea}{\begin{eqnarray}}
\newcommand{\eea}{\end{eqnarray}}

\newcommand{\cA}{{\cal A}}
\newcommand{\cF}{{\cal F}}
\newcommand{\cO}{{\cal O}}
\newcommand{\cB}{{\cal B}}

\newcommand{\cL}{{\cal L}}

\newcommand{\cT}{{\cal T}}
\newcommand{\cM}{{\cal M}}

\newcommand{\SU}{{\rm SU}}
\newcommand{\UU}{{\rm U}}
\newcommand{\heavy}{{[3]}}
\newcommand{\light}{{[12]}}
\newcommand{\bsll}{b\to s\hspace{0.3mm}\ell^+\ell^-}

\newcommand{\no}{\nonumber}

\newcommand{\sla}{\! \! \! \!  /~}
\newcommand{\rhob}{\bar\varrho}
\newcommand{\etab}{\bar\eta}

\renewcommand{\Im}{{\rm Im}\,}
\renewcommand{\Re}{{\rm Re}\,}

\begin{document}

\title{Flavour Physics and CP Violation}

\author{Gino Isidori\thanks{gino.isidori@uzh.ch}}
\affil{Physik-Institut, Universit\"at Z\"urich, 8057 Z\"urich, Switzerland}
\maketitle

\begin{abstract}
These lectures provide a concise introduction to flavor physics, within and beyond the Standard Model,  
with main focus on $B$-physics phenomenology and some recent developments. 
The first lecture is an introduction to the flavor sector of the Standard Model. 
The second lecture is devoted to $B$-meson mixing and rare $B$ decays.  The last lecture contains a general discussion
about  flavor physics beyond the Standard Model,  with a highlighting of recent developments related to the idea of flavor deconstruction.
\end{abstract}

{\footnotesize
\tableofcontents 
}

\section{Preface}

The existence of three nearly identical replicas (or flavors) of quarks and leptons, as well as the origin of their different masses, remains one of the fundamental questions in fundamental physics.  Flavor physics try to address this question and,  more broadly, try to understand whether there are new sources of flavor non-universality  beyond those arising from the fermion mass matrices.  To this end, a basic prerequisite is the understanding of the fascinating interplay of strong and weak interactions in low-energy flavor-changing processes, which in itself is an extremely interesting topic. 

The presence of additional sources of flavor non-universality is a natural expectation in any extension of the Standard Model (SM). Their research is an essential and irreplaceable tool for understanding the ultraviolet completion of the SM. 
While direct searches for new particles at high energies provide a direct information on the mass spectrum of the underlying theory, 
 indirect information from   flavor-changing processes provides unique constraints on theory's couplings. To date, no significant deviations from the SM have been observed. Although frustrating to some extent, this information is an essential ingredient for model building.  As I will discuss in this lectures, 
several possibilities remain open. A particularly interesting one is that new physics exists around the TeV scale, as suggested by a natural stabilization of the Higgs sector, but it has a rather non-trivial flavor structure. Interestingly enough, this hypothesis might even explain  some of the intriguing  hints  observed at present in selected low-energy observables. 

One of the most fashinating aspects of flavor physics nowadays is that we can expect a large increase in experimental data on many processes, more than in other areas of high-energy physics.  Knowledge of flavor-changing processes is therefore likely to improve greatly in the coming years, if supported by appropriate improvements also from the theoretical side.  In these lectures, I will illustrate a few examples of these interesting developments, focusing on $B$-physics phenomenology.

Flavor physics is a very rich field, and what I will be able to cover in these lectures is only a minimal part. More extensive studies can be found, for instance, in these recent books~\cite{Grossman:2023wrq,Burasbook,Artuso:2022ijh} and lecture series~\cite{Silvestrini:2019sey,Zupan:2019uoi,Altmannshofer:2024ykf}, 
and in these review articles~\cite{Buchalla:1995vs,Isidori:2010kg,Altmannshofer:2024jyv}. The present three lectures are organised as follows: in the first lecture I recall the main features of  flavor physics within the SM.  The second lecture is devoted to specific aspects of $B$-physics phenomenology, in particular $B$-meson mixing and rare $B$ decays.  In the last lecture I discuss flavor physics beyond the SM, both from a general effective theory point of view and in the specific  case of flavor non-universal gauge theories, which represent one of the most interesting recent developments in this field.

%\newpage 

\section{Lecture I: The  flavor sector of the SM}

The Standard Model (SM) Lagrangian can be divided into two main parts, 
the gauge and the Higgs (or symmetry breaking) sector. The gauge sector
is extremely simple and highly symmetric: it is completely specified by the 
{\em local} symmetry ${\mathcal G}^{\rm SM}_{\rm local} =SU(3)_{C}\times SU(2)_{L}\times U(1)_{Y}$
and by the fermion content,
\bea
\cL^{\rm SM}_{\rm gauge} &=& \sum_{i=1\ldots3}\ \sum_{\psi=Q^i_L \ldots E^i_R}
{\bar \psi} i D\sla \psi \no\\
&& -\frac{1}{4} \sum_{a=1\ldots8} G^a_{\mu\nu} G^a_{\mu\nu} -\frac{1}{4}
 \sum_{a=1\ldots3} W^a_{\mu\nu} W^a_{\mu\nu}  -\frac{1}{4} B_{\mu\nu} B_{\mu\nu}~.
\eea
The fermion content consist of five fields with different quantum numbers 
under the gauge group.\footnote{~The notation used to indicate each field
is $\psi(A,B)_Y$, where $A$ and $B$ denote the representation under the 
$SU(3)_{C}$ and $SU(2)_L$ groups, respectively,  and $Y$ is the $U(1)_Y$ charge.}
\be
\label{eq:SMfer}
Q^i_{L}(3,2)_{+1/6}~,\ \ U^i_{R}(3,1)_{+2/3}~,\ \
D^i_{R}(3,1)_{-1/3}~,\ \ L^i_{L}(1,2)_{-1/2}~,\ \ E^i_{R}(1,1)_{-1}~,
\ee
each of them appearing in three different replica or 
flavors ($i=1,2,3$).

The threefold structure of fermionic fields gives rise to a large {\em global} flavor symmetry.
Both the local and the global symmetries of $\cL^{\rm SM}_{\rm gauge}$
are broken by the Higgs sector, i.e.~by the part of the Lagrangian describing the dynamics 
of the scalar field $H$  transforming as a doublet of  $SU(2)_L$,
 \be
H = \left(  \ba{c} \phi^+  \\ \phi^0 \ea \right)~.
\ee
The Higgs potential is such that $H$ acquires a non-vanishing  vacuum expectation value (vev)
and this leads to the  {\em spontaneous breaking} of the gauge symmetry. 
The Higgs vev assumes the form
\be
\langle 0 | H | 0 \rangle  = \frac{1}{\sqrt{2}} \left(  \ba{c} 0  \\ v \ea \right)~,
\ee
where $v$ can be determined by the $W$ boson mass
\be
m_W^2 = \frac{g^2 v^2}{4}~, \qquad 
v  = (\sqrt{2} G_F)^{-1/2} \approx 246~{\rm GeV}~.
\ee
The global flavor symmetry is {\em explicitly broken} by 
the Yukawa interaction of $H$ with the fermion fields:
\be
\label{eq:SMY}
- {\cal L}^{\rm SM}_{\rm Yukawa}=Y_d^{ij} {\bar Q}^i_{L} D^j_{R} H
 +Y_u^{ij} {\bar Q}^i_{L}  U^j_{R} H_c + Y_e^{ij} {\bar L}_{L}^i
E_{R}^j H  + {\rm h.c.} \qquad  ( H_c =i\sigma_2 H^\dagger)~.  
\ee

\medskip

The large global flavor symmetry of  $\cL^{\rm SM}_{\rm gauge}$ 
is $U(3)^5$ group, corresponding to the independent unitary rotations in flavor space 
of the five fermion fields in Eq.~(\ref{eq:SMfer}).
This can be decomposed as follows: 
\be 
{\mathcal G}_{\rm flavor} = U(3)^5 = U(1)^5 \times  
{\mathcal G}_{q} \times {\mathcal G}_{\ell}~, 
\label{eq:Gtot}
\ee
where 
\be
{\mathcal G}_{q} = {SU}(3)_{Q_L}\times {SU}(3)_{U_R} \times {SU}(3)_{D_R}, \qquad 
{\mathcal G}_{\ell} =  {SU}(3)_{L_L} \otimes {SU}(3)_{E_R}~.
\label{eq:Ggroups}
\ee
Three of the five $U(1)$ subgroups can be identified with the total barion, the total lepton number,
and the weak hypercharge. None of them is broken by $\cL_{\rm Yukawa}$. 
However, the weak hypercharge,  which is gauged,  is broken spontaneously by $\langle H \rangle  \not=0$. 

The subgroups controlling flavor-changing dynamics and flavor non-universality  are the non-Abelian groups ${\mathcal G}_{q}$ 
and ${\mathcal G}_{\ell}$, which are explicitly broken by $Y_{d,u,e}$.  The structure of these couplings is such that the 
breaking is complete in the quark sector, while two $U(1)$ groups, related to the individual lepton numbers, survive in the lepton sector:
\bea
{\mathcal G}_{q}  	& 	\stackrel{Y_u, Y_d}{\longrightarrow}  & 0\,,   \nonumber \\
{\mathcal G}_{\ell} 	& 	\stackrel{Y_e }{\longrightarrow}  &  U(1)_{L_e - L_\mu} \times     U(1)_{L_\mu  - L_\tau}\,. 
\label{eq:YbreaK}
\eea
This is the maximal breaking possible given the form of the SM Yukawa Lagrangian in (\ref{eq:SMY}). 
Vanishing eigenvalues in  $Y_{d,u,e}$ or a symmetric form for the Yukawa couplings
(such as the identity matrix) would have left larger unbroken subgroups. This does not happen; however, as we shall discuss in more detail later on, 
many of the entries in  $Y_{d,u,e}$ are very small, giving rise to a series of important properties related to the {\em approximate} 
symmetries that hold in the limit when these entries are set to zero.

The diagonalization of each Yukawa coupling requires, in general, two 
independent unitary matrices, $V_L Y V^\dagger_R = {\rm diag}(y_1,y_2,y_3)$.
In the lepton sector the invariance of $\cL^{\rm SM}_{\rm gauge}$ 
under ${\mathcal G}_{\ell}$ allows us to freely choose the two matrices 
necessary to diagonalize  $Y_e$ without breaking gauge invariance, 
or without observable consequences. This is why the three individual lepton numbers 
are conserved, no matter which is the form of $Y_e$. 

The situation is different in the quark 
sector, where we can freely choose only three of the four unitary matrices 
necessary to diagonalize both $Y_{d}$ and $Y_u$. Choosing the basis where 
$Y_{d}$ is diagonal (and eliminating the right-handed 
diagonalization matrix of $Y_u$) we can write 
\be
\label{eq:Ydbasis}
Y_d=\lambda_d~, \qquad  Y_u=V^\dagger\lambda_u~,
\ee
where 
\be
\label{eq:deflamd}
\lambda_d={\rm diag}(y_d,y_s,y_b)~, \ \ \
\lambda_u={\rm diag}(y_u,y_c,y_t)~, \qquad y_q = \frac{m_q}{v}~.
\ee
Alternatively we could choose a  gauge-invariant basis where 
$Y_d= V \lambda_d$ and $Y_u=\lambda_u$. Since the flavor symmetry  
does not allow the diagonalization from the left of both $Y_{d}$ and $Y_u$,
in both cases we are left with a non-trivial unitary mixing matrix, $V$, 
which is nothing but the Cabibbo-Kobayashi-Maskawa (CKM) 
mixing matrix~\cite{Cabibbo:1963yz,Kobayashi:1973fv}.

A generic unitary $3\times3$ [$N\times N$] complex unitary matrix depends 
on three [$N(N-1)/2$] real rotational angles and 
six [$N(N+1)/2$] complex phases. Having chosen a quark basis where 
the $Y_{d}$ and $Y_u$ have the form in (\ref{eq:Ydbasis})
leaves us with a  residual invariance under the flavor group
which allows us to eliminate five of the six complex phases in $V$ 
(the relative phases of the various quark fields).
As a result, the physical parameters in $V$ are four: three real angles and
one complex CP-violating phase. 
The full set of parameters controlling 
the breaking of the quark flavor symmetry in the SM is composed by the 
six quark masses in $\lambda_{u,d}$ and the four parameters of $V$.

For practical purposes it is often convenient to work in the mass eigenstate basis 
of both up- and and down-type quarks. This can be achieved rotating independently 
the up and down components of the quark doublets 
\be
Q^i_L = \left(  \ba{c}  u^i_L     \\  d^i_L  \ea \right)~.
\ee
By these procedure, we  ``move''  the CKM matrix 
from the Yukawa sector to the charged weak current in $\cL^{\rm SM}_{\rm gauge}$:
\be
\left. J_W^\mu \right|_{\rm quarks} = \bar u^i_L \gamma^\mu d^i_L \quad \stackrel{u,d~{\rm mass-basis}}{\longrightarrow} \quad
\bar u^i_L V_{ij} \gamma^\mu d^j_L ~.
\label{eq:Wcurrent}
\ee
However, it must be stressed that $V$ originates from the Yukawa sector (in particular 
by the miss-alignment of $Y_u$ and $Y_d$ in the ${SU}(3)_{Q_L}$ subgroup of ${\mathcal G}_q$): 
in absence of Yukawa  couplings we can always set $V_{ij}=\delta_{ij}$.

\subsection{Properties of the CKM matrix}

The standard parametrization of the CKM matrix~\cite{Chau:1984fp}
in terms of three rotational angles ($\theta_{ij}$) 
and one complex phase ($\delta$) is 
\bea
V&=&\left(\begin{array}{ccc}
V_{ud}&V_{us}&V_{ub}\\
V_{cd}&V_{cs}&V_{cb}\\
V_{td}&V_{ts}&V_{tb}
\end{array}\right)  = R(s_{12}) \times R(s_{13}; e^{i\delta} ) \times R(s_{23}) \no \\
&=&
\left(\begin{array}{ccc}
c_{12}c_{13}&s_{12}c_{13}&s_{13}e^{-i\delta}\\ -s_{12}c_{23}
-c_{12}s_{23}s_{13}e^{i\delta}&c_{12}c_{23}-s_{12}s_{23}s_{13}e^{i\delta}&
s_{23}c_{13}\\ s_{12}s_{23}-c_{12}c_{23}s_{13}e^{i\delta}&-s_{23}c_{12}
-s_{12}c_{23}s_{13}e^{i\delta}&c_{23}c_{13}
\end{array}\right)~,
\label{eq:Chau}
\eea
where $c_{ij}=\cos\theta_{ij}$, $s_{ij}=\sin\theta_{ij}$,
$R(\theta_{12})$ and $R(\theta_{23})$
are real $2\times 2$ rotational matrices (among the 1--2  and 2--3 families, respectively),
and $R(s_{13}; e^{i\delta})$ is 
\be
R(s_{13}; e^{i\delta} ) \left(\begin{array}{ccc}
c_{13}& 0 & s_{13}e^{-i\delta}\\0 & 1 & 0 \\ -s_{13}e^{i\delta}& 0 & c_{13}
\end{array}\right)~.
\ee

Under the phase field redefintions
$u^i_L \to e^{i\alpha^{u}_i} u^i_L$ and 
$d^i_L \to e^{i\alpha^{d}_i} d^i_L$, the CKM elements 
are transformed as 
\be
V_{ij} \to e^{i(\alpha^{d}_j-\alpha^{u}_i)} V_{ij}~. 
\ee
This implies that the moduli of the elements ($|V_{ij}|$) and the combinations 
\be
\frac{V_{ai}^{}V_{aj}^*}{V_{bi}^{}V_{bj}^*}~,
\ee
are phase-convention independent quantities.

The off-diagonal elements of the CKM matrix
show a strongly 
hierarchical pattern:  $|V_{us}|$ and $|V_{cd}|$ are close to $0.22$, the elements
$|V_{cb}|$ and $|V_{ts}|$ are of order $4\times 10^{-2}$ whereas $|V_{ub}|$ and
$|V_{td}|$ are of order $5\times 10^{-3}$. 
The Wolfenstein parametrization, namely the expansion of the CKM matrix 
elements in powers of the small parameter $\lambda \doteq |V_{us}| \approx 0.22$, is a
convenient way to exhibit this hierarchy in a more explicit way~\cite{Wolfenstein:1983yz}:
\begin{equation}
V=
\left(\begin{array}{ccc}
1-{\lambda^2\over 2}&\lambda&A\lambda^3(\varrho-i\eta)\\ -\lambda&
1-{\lambda^2\over 2}&A\lambda^2\\ A\lambda^3(1-\varrho-i\eta)&-A\lambda^2&
1\end{array}\right)
+{\cal{O}}(\lambda^4)~,
\label{eq:Wolfpar} 
\end{equation}
where $A$, $\varrho$, and $\eta$ are free parameters of order 1. 
Because of the smallness of $\lambda$ and the fact that for each element 
the expansion parameter is actually $\lambda^2$, this is a rapidly converging
expansion.

The Wolfenstein parametrization is certainly more transparent than
the standard para\-metrization. However, if one requires sufficient 
level of accuracy, the terms of ${\cal{O}}(\lambda^4)$ and 
${\cal{O}}(\lambda^5)$ have to be included in phenomenological applications.
This can be achieved in many different ways, according to the convention 
adopted. The simplest (and nowadays commonly adopted) choice is obtained 
{\it defining} the parameters $\{\lambda,A,\varrho,\eta\}$ in terms of 
the angles of the exact parametrization in Eq.~(\ref{eq:Chau}) as follows~\cite{Buras:1994ec}:
\begin{equation}
\label{eq:rhodef} 
\lambda\doteq s_{12}~,
\qquad
A \lambda^2\doteq s_{23}~,
\qquad
A \lambda^3 (\varrho-i \eta)\doteq s_{13} e^{-i\delta}~.
\end{equation}
% which implies 
% \begin{equation}\label{2.84} 
% \varrho=\frac{s_{13}}{s_{12}s_{23}}\cos\delta~,
% \qquad
% \eta=\frac{s_{13}}{s_{12}s_{23}}\sin\delta~.
% \end{equation}
The change of variables $\{ s_{ij}, \delta \} \to \{\lambda,A,\varrho,\eta\}$
in Eq.~(\ref{eq:Chau}) leads to an exact parametrization 
of the CKM matrix in terms of the Wolfenstein parameters.  
Expanding this expression up to ${\cal{O}}(\lambda^5)$ leads to
\begin{equation}
\label{eq:Buraspar} 
\left(\begin{array}{ccc}
1-\frac{1}{2}\lambda^2-\frac{1}{8}\lambda^4               &
\lambda+{\cal{O}}(\lambda^7)                                   & 
A \lambda^3 (\varrho-i \eta)                              \\
-\lambda+\frac{1}{2} A^2\lambda^5 [1-2 (\varrho+i \eta)]  &
1-\frac{1}{2}\lambda^2-\frac{1}{8}\lambda^4(1+4 A^2)     &
A\lambda^2+{\cal{O}}(\lambda^8)                                \\
A\lambda^3(1-\rhob-i\etab)                       &  
\!\!\!\!\!  -A\lambda^2+\frac{1}{2}A\lambda^4[1-2 (\varrho+i\eta)]   &
1-\frac{1}{2} A^2\lambda^4                           
\end{array}\right)
\end{equation}
where
\begin{equation}
\label{eq:rhobar}
\rhob = \varrho (1-\frac{\lambda^2}{2})+{\cal O}(\lambda^4)~,
\qquad
\etab=\eta (1-\frac{\lambda^2}{2})+{\cal O}(\lambda^4)~.
\end{equation}
The advantage of this generalization of the Wolfenstein parametrization
is the absence of relevant corrections to $V_{us}$, $V_{cd}$, $V_{ub}$ and 
$V_{cb}$, and a simple change in $V_{td}$, which facilitate the implementation 
of experimental constraints. 

\begin{figure}[t]
\begin{center}
\includegraphics[width=8cm]{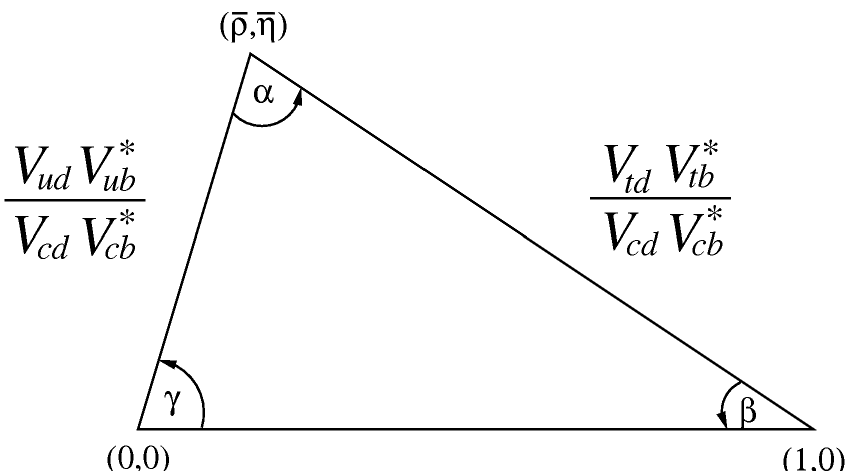}
\end{center}
\caption{The CKM unitarity triangle.}
\label{fig:1utriangle}
\end{figure}

The unitarity of the CKM matrix implies the following relations between its
elements:
\be
{\bf I)}\quad 
 \sum_{k=1\ldots 3} V_{ik}^* V_{ki}=1~,
\quad\qquad 
{\bf II)}\quad
\sum_{k=1\ldots 3} V_{ik}^* V_{kj\not=i}~.
\ee
These relations are a distinctive feature of the SM, where the CKM matrix is the only 
source of quark flavor mixing.  Their experimental verification is therefore a useful 
tool to set bounds, or possibly reveal, new sources of flavor symmetry breaking. 
Among the relations of type {\bf II}, the one obtained for $i=1$ and $j=3$, 
namely 
\be
V_{ud}^{}V_{ub}^* + V_{cd}^{}V_{cb}^* + V_{td}^{}V_{tb}^* =0 
\label{eq:UT}
\ee
\be
{\rm or} \qquad 
\frac{V_{ud}^{}V_{ub}^*}{V_{cd}^{}V_{cb}^*}  + \frac{V_{td}^{}V_{tb}^*}{V_{cd}^{}V_{cb}^*}  + 1 = 0
\qquad \leftrightarrow \qquad
 [\rhob+i \etab] + [(1-\rhob)-i\etab] + 1 =0~,
\no
\ee
is particularly interesting since it involves the sum of three terms all
of the same order in $\lambda$ and is usually represented as a unitarity triangle
in the complex  plane, as shown in Fig.~\ref{fig:1utriangle}.
It is worth to stress that Eq.~(\ref{eq:UT}) is invariant under any 
phase transformation of the quark fields. Under such transformations
the triangle in Fig.~\ref{fig:1utriangle} is rotated in the complex plane,
but its angles and the sides remain unchanged.
Both angles and  sides of the unitary triangle are indeed observable quantities
which can be measured in suitable experiments.

\begin{figure}[t]
  \centering
\raisebox{-7pt}{\includegraphics[width=0.46\textwidth,height=0.46\textwidth]{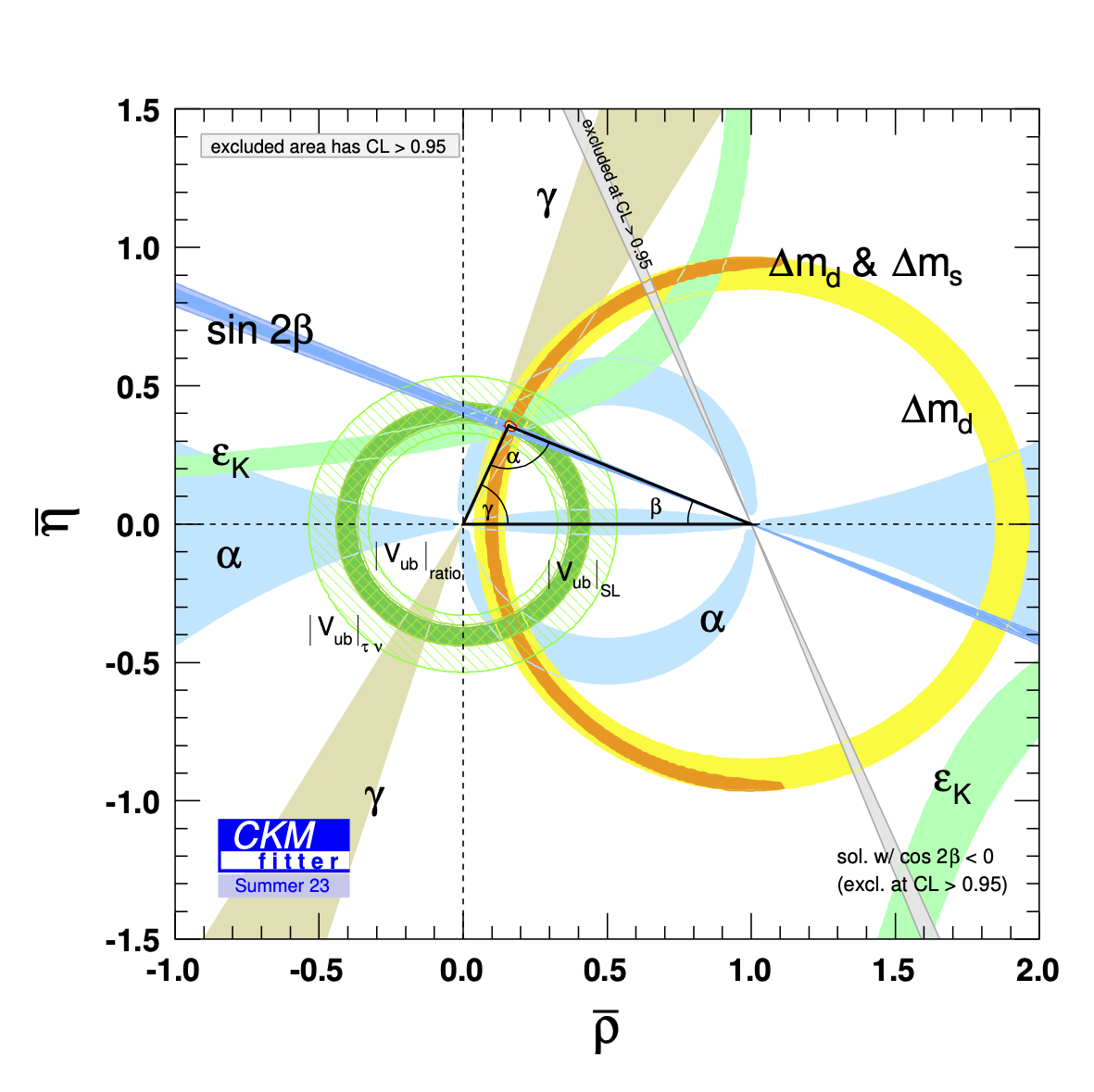}}
\includegraphics[width=0.52\textwidth,height=0.44\textwidth]{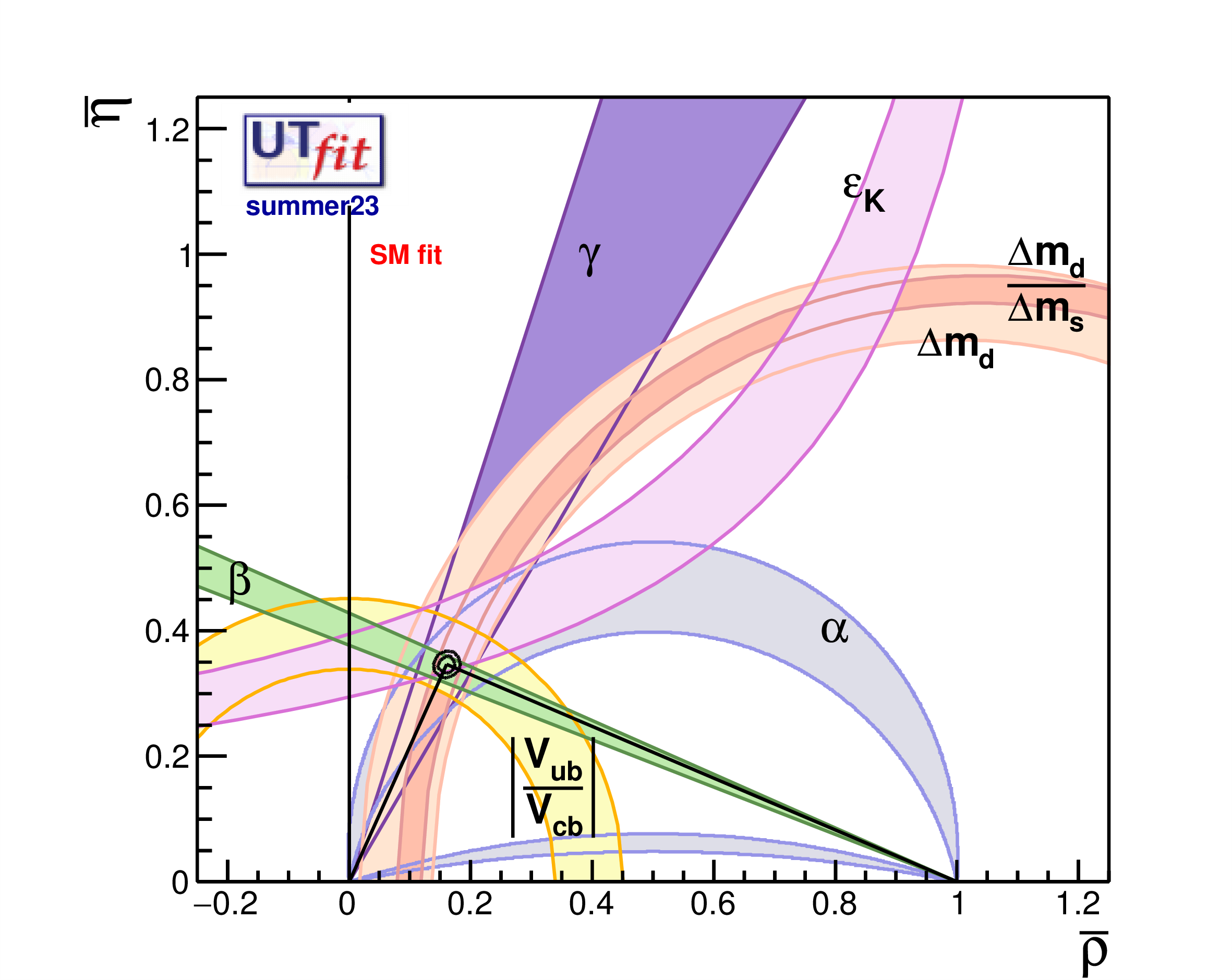}
  \caption{Allowed region in the $\rhob,\etab$ plane, from the CKMfitter~\cite{ValeSilva:2024jml} (left) and  the UTfit~\cite{Bona:2024bue} (right)
  collaborations.  Superimposed are
  the individual constraints from charmless semileptonic $B$ decays
  ($|V_{ub}|$), mass differences in the $B_d$ ($\Delta m_d$)
  and $B_s$ ($\Delta m_s$) systems, CP violation 
  in the neutral kaon system ($\varepsilon_K$) and in the $B_d$ systems ($\sin2\beta$),
  the combined constrains on $\alpha$ and $\gamma$ from various $B$ decays. }
  \label{fig:UT}
\end{figure}

\medskip

The values of $|V_{us}|$ and $|V_{cb}|$, or $\lambda$ and $A$ in the parametrization 
(\ref{eq:Buraspar}), are determined with good accuracy from  $K\to\pi\ell\nu$ and 
$B\to X_c \ell\nu$ decays, respectively. According to recent phenomenological 
analyses~\cite{Bona:2024bue}  their numerical values are
\be
\lambda=0.2251  \pm0.0008~,\qquad A=0.83\pm0.01~.
\ee
Using these results, all the other constraints on the elements of 
the CKM matrix can be expressed as constraints on $\rhob$ and $\etab$
(or constraints on the CKM unitarity triangle in Fig.~\ref{fig:1utriangle}).
The resulting constraints are shown in Fig.~\ref{fig:UT}.
As can be seen, they are all consistent with a unique value of 
$\rhob$ and $\etab$.

The consistency of different constraints 
on the CKM unitarity triangle is a powerful consistency test 
of the SM in describing flavor-changing phenomena.
From the plot in Fig.~\ref{fig:UT} it is quite clear, 
at least in a qualitative way, that there is little room 
for non-SM contributions in flavor changing transitions.

\subsection{Low-energy effective Lagrangians}
\label{sect:Heff}

The decays of $B$, $D$, and $K$ 
mesons are processes which involve at least two different 
energy scales: the electroweak scale, characterized by the $W$ boson mass, 
which determines the flavor-changing transition at the quark level, 
and the scale of strong interactions $\Lambda_\mathrm{QCD}$,
related to the hadron formation. The presence of these two widely separated 
scales makes the calculation of the decay amplitudes starting from the 
full SM Lagrangian quite complicated: large logarithms of the type 
log($m_W/\Lambda_\mathrm{QCD}$) may appear, leading to a breakdown of 
ordinary perturbation theory. 

This problem can be substantially simplified by integrating out the 
heavy SM fields ($W$ and $Z$ bosons, as well as the top quark)
at the electroweak scale, and constructing an appropriate low-energy 
effective field theory (EFT) where only the light SM fields appear.
The weak effective Lagrangians thus obtained 
contains local operators of dimension six (and higher), written 
in terms of light SM fermions, photon and gluon fields, suppressed 
by inverse powers of the $W$ mass. 

To be concrete, let's consider the example of charged-current 
semileptonic weak interactions. The basic building 
block in the full SM Lagrangian is
\be
\cL^{\rm full~SM}_{W} = \frac{g}{\sqrt{2}} J_{W}^\mu(x) W_\mu^+(x) + {\rm h.c.}~,
\ee
where 
\be
J_W^\mu(x)= V_{i j}~\bar u_L^i(x)\gamma^\mu d_L^j(x)
+\bar e^j_L(x) \gamma^\mu \nu_L^j(x) 
\ee
is the weak charged current already introduced in Eq.~(\ref{eq:Wcurrent}).
Integrating out the $W$ field at the tree level we contract
two vertexes of this type generating the  non-local transition amplitude
%\footnote{To be more precise, this amplitude is the 
%~We recall that, given a generic
%Lagrangian $\cL(\phi(x))$, the $S$ matrix can be written as the 
%functional integral $\int \cD\phi~{\rm exp}\left[ i \int d^4x \cL(\phi(x)) \right]$.}
\be
i \cT = - i \frac{g^2}{2} \int d^4x D_{\mu \nu} \left( x, m_W \right) 
T \left[ J_W^\mu(x),J_W^{\nu\dagger}(0) \right]~,
\ee
which involves only light fields. Here $D_{\mu \nu} \left( x, m_W \right)$ is 
the $W$ propagator in coordinate space: expanding it in inverse powers of $m_W$,
\begin{equation}
D_{\mu \nu} \left( x, m_W \right)=\int \frac{d^4 q}{(2\pi)^4} e^{-i q\cdot x} \frac{-i g_{\mu\nu} + \cO(q_\mu,q_\nu) }
{q^2-m_W^2+i\varepsilon}=\delta(x) \frac{ i g_{\mu\nu}}{m_W^2}+\dots\,,
\label{eq:MWexp}
\end{equation}
the leading contribution to $\cT$  can be interpreted as  
the tree-level contribution of the following effective local Lagrangian 
\be
\cL^{\rm (0)}_{\rm eff} = - \frac{4 G_F}{\sqrt{2}}  g_{\mu\nu} 
J_W^\mu(x) J_W^{\nu\dagger}(x)~,
\label{eq:LFermi0}
\ee
where  $G_F/\sqrt{2}=g^2/(8 m_W^2)$ is the Fermi coupling.
If we select in the product of the two currents
one quark and one leptonic current, 
\be
\cL^{\rm semi-lept}_{\rm eff} = - \frac{4 G_F}{\sqrt{2}}~V_{i j}~\bar u_L^i(x)\gamma^\mu d_L^j(x)
~\bar \nu_L(x) \gamma_\mu e_L(x) + {\rm h.c.}~,
\label{eq:LFermi}
\ee
we obtain an effective Lagrangian 
which provides an excellent description of semileptonic weak decays.
For $B$ decays the neglected terms in the expansion (\ref{eq:MWexp}) correspond
to corrections of $\cO(m_B^2/m^2_W)$ to the decay amplitudes. In principle,
these corrections could be taken into account by adding appropriate 
dimension-eight operators in the effective Lagrangian. However, in most 
cases they are safely negligible.

The case of charged semileptonic decays is particularly simple since we can 
ignore QCD effects: the operator (\ref{eq:LFermi}) is not renormalized 
by strong interactions (a more detailed explanation of why this happens will
be presented in the next lecture). This is the main reason we can determine 
with high precision a few moduli of CKM matrix elements (in particular 
$|V_{cb}|$ and $|V_{us}|$) from semileptonic decays of $B$ and $K$ mesons.

The situation is more 
complicated in four-quark interactions and flavor-changing neutral-current
processes, where QCD corrections and higher-order weak interactions 
cannot be neglected, but the basic strategy is the same. 
First of all we need to identify a complete basis of local operators, 
that includes also those generated beyond the tree level. In general,
given a fixed order in the  $1/m^2_W$ expansion of the amplitudes,
we need to consider all operators of corresponding dimension 
(e.g.~dimension six at the first order in the  $1/m^2_W$ expansion)
compatible with the symmetries of the system. Then we must introduce 
an artificial scale in the problem, the renormalization scale $\mu$,
which is needed to regularize QCD (or QED) corrections 
in the EFT.

The effective Lagrangian for generic $\Delta F=1$ processes\footnote{We denote by 
$\Delta F=1$ processes, the transitions with change of flavor by one ``unit'', such as 
e.g.~$b\to s \bar u u$. Here we the initial state has $b$-flavor=1 and $s$-flavor=0, whereas the 
final state has  $b$-flavor=0 and $s$-flavor=1, hence $\Delta F_b = -\Delta F_s =1$.} 
assumes the form
\be
\cL_{\Delta F=1} =  - 4\frac{G_F}{\sqrt{2}}  \sum_i C_i (\mu)  Q_i 
\label{eq:effH} 
\ee
where the sum runs over the complete basis of operators.
As explicitly indicated, the effective couplings $C_i(\mu)$
(known as Wilson coefficients) depend, in general,
on the renormalization scale, similarly to the 
scale dependence of the QCD coupling $\alpha_s(\mu)$.
The dependence from this
scale cancels when evaluating the matrix elements of the effective 
Lagrangian for physical processes, that we 
can generically indicate as 
\be
\cM (i\to f) =  - 4\frac{G_F}{\sqrt{2}}  \sum_i C_i (\mu) \langle f |  Q_i(\mu) | i \rangle~.
\ee
The scale $\mu$ acts as a separator of short- and long-distance virtual 
corrections: short-distance effects are included in the  $C_i(\mu)$,
whereas long-distance effects are left as explicit degrees
of freedom in the EFT. 

In practice, the problem reduces to the following three 
well-defined and independent steps:
\begin{enumerate}
\item[1.] the evaluation of the {\em initial conditions} of the 
$C_i (\mu)$ at the electroweak scale $(\mu \approx m_W)$, that 
is done by an approrpiate {\em matching procedure} between the
effective theory and the full theory;
\item[2.] the evaluation of the {\em renormalization-group equations} (RGE)
which determine the {\em evolution} 
of the $C_i (\mu)$ from the electroweak scale down 
to the energy scale of the physical process ($\mu \approx m_B$);
\item[3.] the evaluation of the {\em matrix elements} of the 
effective Lagrangian for the physical hadronic processes
(which involve energy scales from the meson masses down to $\Lambda_{QCD}$).
\end{enumerate}

The first step is the one where New Physics (NP), 
namely physics beyond the SM, may contribute:
if we assume NP is heavy, it may modify the initial 
conditions of the Wilson coefficients at the high scale, 
while it cannot affect the following two steps.
While the RGE evolution and the hadronic matrix elements are not 
directly related to NP, they may influence the sensitivity to NP 
of physical observables. In particular, the evaluation of 
hadronic matrix elements is potentially affected by non-perturbative 
QCD effects: these are often a large source of theoretical uncertainty 
which can obscure NP effects. RGE effects do not induce sizable uncertainties
since they can be fully handled within perturbative QCD;
however, the sizable logs generated by the RGE running may ``dilute''
the interesting short-distance information encoded in the 
value of the Wilson coefficients at the high scale.
As we will discuss in the following, only in specific 
observables these two effects are small and under good theoretical control.

\subsection{Effective Lagrangian for $\Delta F=2$ amplitudes}
\label{sect:DF2}

Among four-quark interactions, a particularly interesting case is those of 
 $\Delta F=2$ transitions, namely amplitudes with a double change 
 of flavor (e.g.~$\Delta F_b = -\Delta F_s =2$). The amplitudes control the 
 mixing of neutral mesons (e.g.~$B^0_s$--$\bar B^0_s$ mixing).
 
\begin{figure}[t]
  \centering
  {\includegraphics[width=0.6\textwidth]{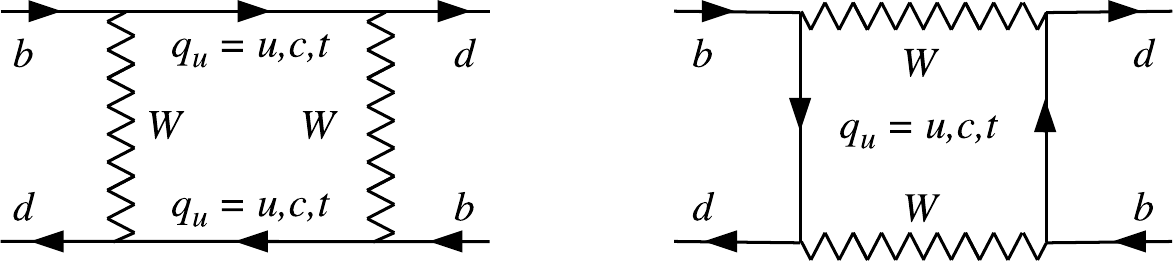}}
 \caption{Box diagrams contributing to $B_d$-$\bar B_d$ mixing in the unitary gauge.}
  \label{fig:BBmix}
\end{figure}

The effective Lagrangian relevant for 
$B^0_d$--$\bar B^0_d$ and $B^0_s$--$\bar B^0_s$ mixing can be 
conventionally written as 
\begin{equation}
{\cal L}^{\rm SM}_{\Delta B=2} = \sum_{q=d,s} C^{\Delta F=2}_q~(\bar b_L \gamma_\mu q_L\, \bar b_L \gamma^\mu q_L)~,
\label{eq:db2}
\end{equation}
where the leading contribution to the Wilson coefficient $C^{\Delta F=2}_q$ 
can be determined by computing the box diagrams in Fig.~\ref{fig:BBmix}
(in the limit of small external momenta). 
The explicit calculation for the coefficient $C^{\Delta F=2}_d$ yields 
\bea
\left. C^{\Delta F=2}_d \right|_{\rm 1-loop} &=& \sum_{q^\prime =u,c,t}  \sum_{q =u,c,t}   (V_{q^\prime b}^*V_{ q^\prime d}) 
 (V_{q b}^*V_{ q d})  F(x_{q^\prime}, x_{q}) \label{eq:DF2} \\ 
 & \approx &  (V_{t b}^*V_{ t d})^2 \left[ F(x_t, x_t) + F(0,0) - 2 F(0,x_t) \right]   \\ 
&= & (V_{tb}^*V_{tq})^2 ~ \frac{G_F^2}{4\pi^2}m_W^2 ~S_0(x_t)
\eea
where $x_q =m_q^2/m_W^2$. The intermediate result in (\ref{eq:DF2}) follows from 
neglecting all quark masses but for $m_t$,  and using the unitarily of the CM matrix 
(that implies $V_{u b}^*V_{ u d} + V_{c b}^*V_{ c d}+V_{t b}^*V_{ t d}=0$).
The latter implies in particular that the mass-independent contribution to the amplitude 
of up, charm, and top-quarks cancel. This cancellation, known as GIM~mechanism, 
was identified for the first time by  Glashow, Iliopoulos and Maiani in 1970, 
when postulating the existence of the  charm quark~\cite{Glashow:1970gm}.
The explicit expression of the loop function is 
\be
S_0(x_t) =  
\frac{4x_t-11x_t^2+x_t^3}{4(1-x_t)^2} -\frac{3x_t^3 \ln x_t}{2(1-x_t)^3}
\label{eq:S0}
\ee  
and, as expected, vanishes in the limit $x_t \to 0$.

\subsection{The gaugeless limit of  $\Delta F=2$ amplitudes}

\begin{figure}[t]
\vskip 0.5 cm
  \centering
  {\includegraphics[width=0.6\textwidth]{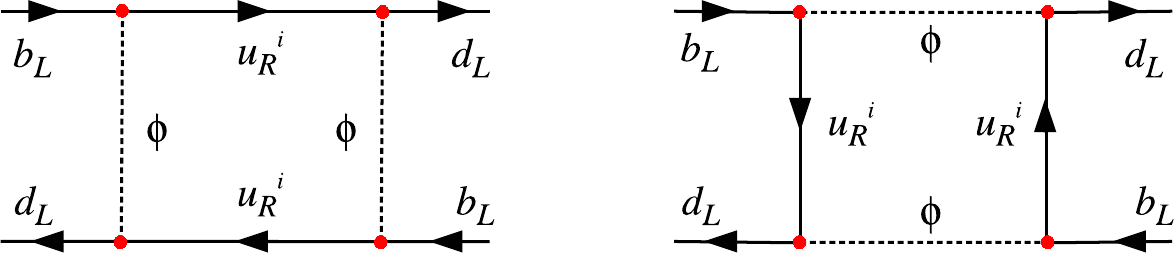}}
\vskip 0.5 cm
 \caption{Box diagrams contributing to  $B_d$-$\bar B_d$ 
in the gaugeless limit.}
  \label{fig:Gaugeless}
\end{figure}

An interesting aspect of the  loop
function in Eq.~(\ref{eq:S0}) is the fact that 
it diverges in the limit $m_t/m_W \to \infty$. This behavior is 
apparently strange: it contradicts the expectation that 
contributions of heavy particles 
decouple, at low energies, 
in the limit where their masses  increase. 
The origin of this effect can be understand by noting that the 
leading contribution to the $\Delta F=2$ amplitude is generated only by the Yukawa
interaction. This leading contribution can be better isolated in the {\em gaugeless} 
limit of the SM, i.e.~if we send to zero the gauge couplings.
In this limit $m_W\to 0$ and the derivation of the effective 
Lagrangian discussed in Sect.~\ref{sect:Heff} does not make sense.
However, the leading contribution to the effective 
Lagrangians for $\Delta F=2$ processes remains unaffected.
Indeed, the leading contribution to these processes is 
generated by Yukawa interactions of the type in Fig.~\ref{fig:Gaugeless}, 
where the scalar fields are the Goldstone-bosons components of the 
Higgs field (which are not eaten up by the $W$ in the limit $g\to0$).
Since the top is still heavy, we can integrate it out (i.e.~we can compute the 
amplitude in the limit of a heavy top), obtaining
the following result for ${\cal L}_{\Delta B=2}$:
\be
\left. {\cal L}^{\rm SM}_{\Delta B=2} \right|_{g_i \to 0} 
~ = ~
\frac{[(Y_u Y^*_u)_{bq}]^2}{128 \pi^2 m_t^2}  (\bar b_L \gamma_\mu q_L)^2 
~=~ \frac{G_F^2 m_t^2}{16 \pi^2} 
 (V_{tb}^*V_{tq})^2 (\bar b_L \gamma_\mu q_L)^2 ~.
\label{eq:gaugeless}
\ee
Taking into account that $S_0(x)\to x/4$
for  $x \to \infty$, it is easy to verify that this result is equivalent 
to the one in Eq.~(\ref{eq:S0}) in the large $m_t$ limit.

The last expression in Eq.~(\ref{eq:gaugeless}), which holds in the limit where we
neglect the charm Yukawa coupling, shows that the decoupling of the amplitude 
with the mass of the top is compensated by four powers of the 
top Yukawa coupling at the numerator. The divergence for  $m_t\to \infty$
can thus be understood as the divergence of one of the fundamental 
couplings of the theory.  Note also that in the gaugeless limit there 
is no GIM mechanism. The contributions of the various up-type quarks 
inside the loops do not cancel each other: they are directly weighted 
by the corresponding Yukawa couplings, and this is why the top-quark 
contribution is the dominant one. 

This exercise illustrates the key 
role of the Yukawa coupling in determining the main properties
flavor physics within the SM, as advertised at the beginning of this lecture.
It also illustrates the interplay of flavor and electroweak 
symmetry breaking in determining the structure of 
short-distance dominated flavor-changing processes in the SM.

%\newpage 

\section{Lecture II:  $B$-meson mixing and rare $B$ decays}

\subsection{Time evolution and of $B_{d,s}$ states}               
\label{sect:BBmix}

%%%%%%%%%%%%%%%%%%%%%%%%%%%%%%% ++++++++++++++++++++++++++++++ %%%%%%%%%%%%%%%%%%%%%%%%%%%

The non-vanishing amplitude mixing a $B^0$  meson ($B^0_{d}$ or $B^0_{s}$) 
with the corresponding anti meson, described within the SM by the effective Lagrangian
(\ref{eq:db2}), 
induces a time-dependent oscillations between $B^0$ and $\bar B^0$ states:
an initially produced $B^0$ or $\bar B^0$ evolves in time into a
superposition of $B^0$ and $\bar B^0$. Denoting by  $\ket{B^0 (t)} $
(or $\ket{\bar B^0 (t)}$) the
state vector of a $B$ meson which is tagged as a $B^0$ (or $\bar B^0$)
at time $t=0$, the time evolution of these states is governed 
by the following equation:
\be
i \frac{d}{dt} \left(\ba{c} \ket{B(t)} \\ \ket{\bar B(t)} \ea \right)
= \left( M - i\, \frac{\Gamma}{2} \right) 
\left(\ba{c} \ket{B(t)} \\ \ket{ \bar B (t)} \ea \right)~,
\label{mgmat}
\ee
where the mass-matrix $M$ and the decay-matrix $\Gamma$
are $t$-independent, Hermitian $2\times 2$ matrices. 
CPT invariance implies that $M_{11} = M_{22}$ and $\Gamma_{11} = \Gamma_{22}$,
while the off-diagonal element  $M_{12}=M_{21}^*$ is the one we can compute 
using the effective Lagrangian ${\cal L}_{\Delta B=2}$. 

The mass eigenstates are the eigenvectors of $M - i\,\Gamma/2$. We
express them in terms of the flavor eigenstates as
\be
\ket{B_L} = p \ket{B^0} + q \ket{\bar B^0}~,  \qquad 
\ket{B_H} = p \ket{B^0} - q \ket{\bar B^0}~,
\label{defpq}
\ee
with $|p |^2+|q |^2 = 1$.
Note that, in general, $\ket{B_L}$ and $\ket{B_H}$ are not orthogonal 
to each other. The time evolution of the mass eigenstates is 
governed by the two eigenvalues $M_H -i\,\Gamma_H/2$ and $M_L -i\,\Gamma_L/2$:
\be
\ket{B_{H,L} (t) } = e^{-(i M_{H,L} + \Gamma_{H,L}/2)t} \, 
  \ket{B_{H,L}(t=0) } \,. \label{thl}
\ee
For later convenience it is also useful to define 
%\bea
%&& m = \frac{ M_H + M_L}{2} =  M_{11}~,  \qquad \Gamma =  \frac{\Gamma_L + \Gamma_H}2 = \Gamma_{11} \no\\
%&& \Delta m  =  M_H - M_L~, \qquad\qquad\  \Delta \Gamma = \Gamma_L - \Gamma_H~.
%\label{mg} 
%\eea
\be
m = \frac{ M_H + M_L}{2},  \qquad \Gamma =  \frac{\Gamma_L + \Gamma_H}{2}~, \qquad 
\Delta m  =  M_H - M_L~, \qquad  \Delta \Gamma = \Gamma_L - \Gamma_H~.
\label{mg} 
\ee
With these conventions the time evolution of initially tagged $B^0$ or 
$\bar B^0$ states is 
\bea
\ket{B^0 (t)} &=&  e^{-i m t} \, e^{-\Gamma t/2} \left[  
  f_+ (t)\, \ket{B^0} + \frac{q}{p}\, f_- (t)\, \ket{ \bar B^0} \right]~, \no\\
\ket{\bar B^0 (t)} &=& e^{-i m t} \, e^{-\Gamma t/2} \left[ \frac{p}{q}\, f_- (t)\, \ket{B^0} 
  +  f_+(t)\, \ket{\bar B^0} ~\right]~,
  \label{tgg}    
\eea 
where
\bea
f_+ (t) &=& \phantom{-} \cosh\frac{\Delta \Gamma t}{4} \cos\frac{\Delta m t}{2} -   
      i \sinh\frac{\Delta \Gamma t}{4} \sin \frac{\Delta m t}{2}~, \\
%%\qquad \stackrel{\Delta \Gamma=0}{\longrightarrow}  \qquad \cos\frac{\Delta m t}{2}~,  \\
f_- (t) &=&  - \sinh\frac{\Delta \Gamma t}{4} \cos \frac{\Delta m t}{2} +
      i \cosh\frac{\Delta \Gamma t}{4} \sin \frac{\Delta m t}{2}~,  
%% \qquad \stackrel{\Delta \Gamma=0}{\longrightarrow}  \qquad   i \sin \frac{\Delta m t}{2}~.  
\label{gpgm} 
\eea

In both $B_s$ and $B_d$ systems the following hierarchies holds: 
$|\Gamma_{12}| \ll | M_{12}|$ and $\Delta \Gamma \ll \Delta m$.
They are experimentally verified and can be traced back 
to the fact that $|\Gamma_{12}|$ is a genuine long-distance $\cO(G_F^2)$ effect
(it is indeed related to the absorptive part of the box diagrams 
in Fig.~\ref{fig:BBmix}) which do not share the large $m_t$ enhancement of 
 $|M_{12}|$ (which is a short-distance dominated quantity).  
Taking into account this hierarchy leads to the following 
approximate expressions for the quantities appearing in the time-evolution formulae 
in terms of $M_{12}$ and $\Gamma_{12}$:
\bea
\Delta m &=& 2\, |M_{12}| \left[ 1 + 
  {\cal O} \left( \left| \frac{\Gamma_{12}}{M_{12}} \right|^2 \right) \right]~,
  \label{mgsol:a} \\
\Delta \Gamma &=& 2\, |\Gamma_{12}| \cos \phi \left[ 1+  
  {\cal O} \left( \left| \frac{\Gamma_{12}}{M_{12}} \right|^2 \right) \right]~,
  \label{mgsol:b} \\
\frac{q}{p} &=& - e^{- i \phi_B} \left[ 1 - \frac{1}{2}\left| \frac{\Gamma_{12}}{M_{12}} \right|\sin\phi
   + {\cal O} \left( \left| \frac{\Gamma_{12}}{M_{12}} \right|^2 \right) \right]~,
\label{qpsol}  
\eea
where $\phi = {\rm arg}( - M_{12}/\Gamma_{12})$ and $\phi_B$ is the phase of $M_{12}$.
Note that $\phi_B$ thus defined is not measurable and 
depends on the phase convention adopted, while $\phi$ is a phase-convention quantity 
which can be measured in
experiments.
% For most practical purposes
% this phase convention is equivalent to the standard
% phase convention of the CKM matrix. Indeed the leading contribution to  $\Gamma_{12}$ 
% arises by quark-level transitions of the type $b\to c + \bar c s (\bar u d)$,  
% which are real in the Standard CKM phase convention. 

Taking into account the above results, the time-dependent decay rates 
of an initially tagged $B^0$ or $\bar B^0$ state
into some final state $f$ can be written as
\bea
&& \Gamma[B^0(t=0) \to f(t)]  =  {\cal N}_0  | A_f |^2  e^{-\Gamma t}
  \Bigg\{ \frac{1 + | \lambda_f |^2}2 \cosh \frac{\Delta \Gamma  t}{2} \no \\
&& \qquad 
+   \frac{ 1 - | \lambda_f |^2}2 \cos ( \Delta m  t )  
  - \Re \lambda_f  \sinh \frac{\Delta \Gamma  t}{2} 
  - \Im \lambda_f  \sin ( \Delta m  t) \Bigg\} , \no \\
&& \!\!\!\!\! \Gamma[\bar B^0(t=0) \to f(t)] = {\cal N}_0  | A_f |^2  
\left( 1 + \left| \frac{\Gamma_{12}}{M_{12}} \right|\sin\phi  \right) 
  e^{-\Gamma t} \Bigg\{ \frac{1 + | \lambda_f |^2}2 \times \no \\
 && \times  \cosh \frac{\Delta \Gamma  t}{2}  
   - \frac{1 - | \lambda_f |^2}2 \cos ( \Delta m  t )   
    - \Re \lambda_f  \sinh \frac{\Delta \Gamma  t}{2} 
    + \Im \lambda_f  \sin ( \Delta m  t ) \Bigg\}~, \no
\eea
where ${\cal N}_0$ is the flux normalization and, following the standard notation, 
we have defined 
\be
\lambda_f = \frac{q}{p} \frac{\bar A_f}{A_f}
  \approx - e^{- i \phi_B}\, \frac{\bar A_f}{A_f}  \left[ 1 
- \frac{1}{2}\left| \frac{\Gamma_{12}}{M_{12}} \right|\sin\phi \right] 
\ee
in terms of the decay amplitudes 
\be 
A_f = \langle f | \cL_{\Delta F=1} \ket{B^0}~, \qquad \qquad \bar A_f = \langle f |
 \cL_{\Delta F=1} \ket{\bar B^0}~. \label{defaf}
\ee
From the above expressions it is clear that the key 
quantity we can access experimentally in the time-dependent study of $B$ decays 
is the combination $\lambda_f$. Both real and imaginary parts of  $\lambda_f$
can be measured, and indeed this is a phase-convention  independent quantity: the phase 
convention in $\phi_B$ is compensated by the phase convention in the decay
amplitudes. In other words, what we can measure is the weak-phase difference 
between $M_{12}$ and the decay amplitudes. 

\subsubsection*{CP violation in $B_d$ mixing}

For generic final states, $\lambda_f$ is a quantity that is difficult to evaluate.
However, it becomes particularly simple in the case where $f$ is a CP eigenstate,
${\rm CP}\ket{f} = \eta_f \ket{f}$, and the weak phase of the decaying 
amplitude is know. In such case  $\bar A_f/A_f$ is a pure phase factor ($|\bar A_f / A_f|=1$), 
determined by the weak phase of the decaying amplitude:
\be
\left. \lambda_f \right|_{\rm CP-eigen.} 
= \eta_f \frac{q}{p} e^{-2i \phi_{A} }~, \qquad A_f =| A_f| e^{i \phi_A}~, \qquad \eta_f=\pm 1~.
\ee
The most clean example of this type of channels is the $\ket{\psi K_S}$ final state for 
$B_d$ decays. In this case the final state is a CP eigenstate and the decay amplitude 
is real (to a very good approximation) in the standard CKM phase convention. 
Indeed the underlying partonic transition is dominated by the Cabibbo-allowed 
tree-level process $b\to c \bar c s$, which has a vanishing phase in the 
standard CKM phase convention, and also the leading one-loop corrections 
(top-quark penguins) have the same vanishing weak phase.
Since in the $B_d$ system we can safely neglect $\Gamma_{12}/M_{12}$,
this implies
\be
\lambda^{B_b}_{\psi K_s} = - e^{- i \phi_{B_d} }~, 
\qquad \Im\left(\lambda^{B_b}_{\psi K_s}\right)_{\rm SM} = \sin(2\beta) ~,
\ee
where the SM expression of $\phi_{B_d}$ 
is nothing but the phase of the CKM combination $(V_{tb}^* V_{td})^2$ appearing in 
Eq.~(\ref{eq:db2}). Given the smallness of $\Delta \Gamma_d$, 
this quantity is easily extracted by the ratio
\bea
\frac{ \Gamma[\bar B_d(t=0) \to \psi K_s (t)] - \Gamma[B^0(t=0) \to f \psi K_s (t) ] }
{ \Gamma[\bar B_d(t=0) \to \psi K_s (t)] + \Gamma[B^0(t=0) \to f \psi K_s (t) ] }
 \approx  \Im\left(\lambda^{B_b}_{\psi K_s}\right) \sin(\Delta m_{B_d} t)~.
\no
\eea
The determination of $\sin(2\beta)$, within the SM, using this method, 
has been one of the most significant results of $B$ factories: 
this is how the narrow  $\sin(2\beta)$ constraint in 
Fig.~\ref{fig:UT} is obtained.

Another class of interesting final states are CP-conjugate channels $\ket{f}$
and $\ket{\bar f}$ which are accessible only to $B^0$ or $\bar B^0$ states, 
such that $|A_f| = |\bar A_{\bar f}|$ and  $\bar A_f = A_{\bar f}$=0. Typical examples 
of this type are the charged semileptonic channels. In this case the asymmetry 
\bea
\frac{ \Gamma[\bar B^0(t=0) \to  f (t)] - \Gamma[B^0(t=0) \to \bar f (t) ] }{ 
\Gamma[\bar B^0(t=0) \to f (t)] + \Gamma[B^0(t=0) \to \bar f  (t) ] }
= \left|\frac{\Gamma_{12}}{M_{12}}\right| \sin\phi \left[ 1  
+ {\cal O} \left( \frac{\Gamma_{12}}{M_{12}} \right) \right] 
\no
\eea
turns out to be time-independent and a clean way to determine the indirect
CP-violating phase~$\phi$.

\subsubsection*{CP violation  in $B_s$ mixing}

\begin{figure}[t]
  \centering
  \includegraphics[width=0.6\textwidth]{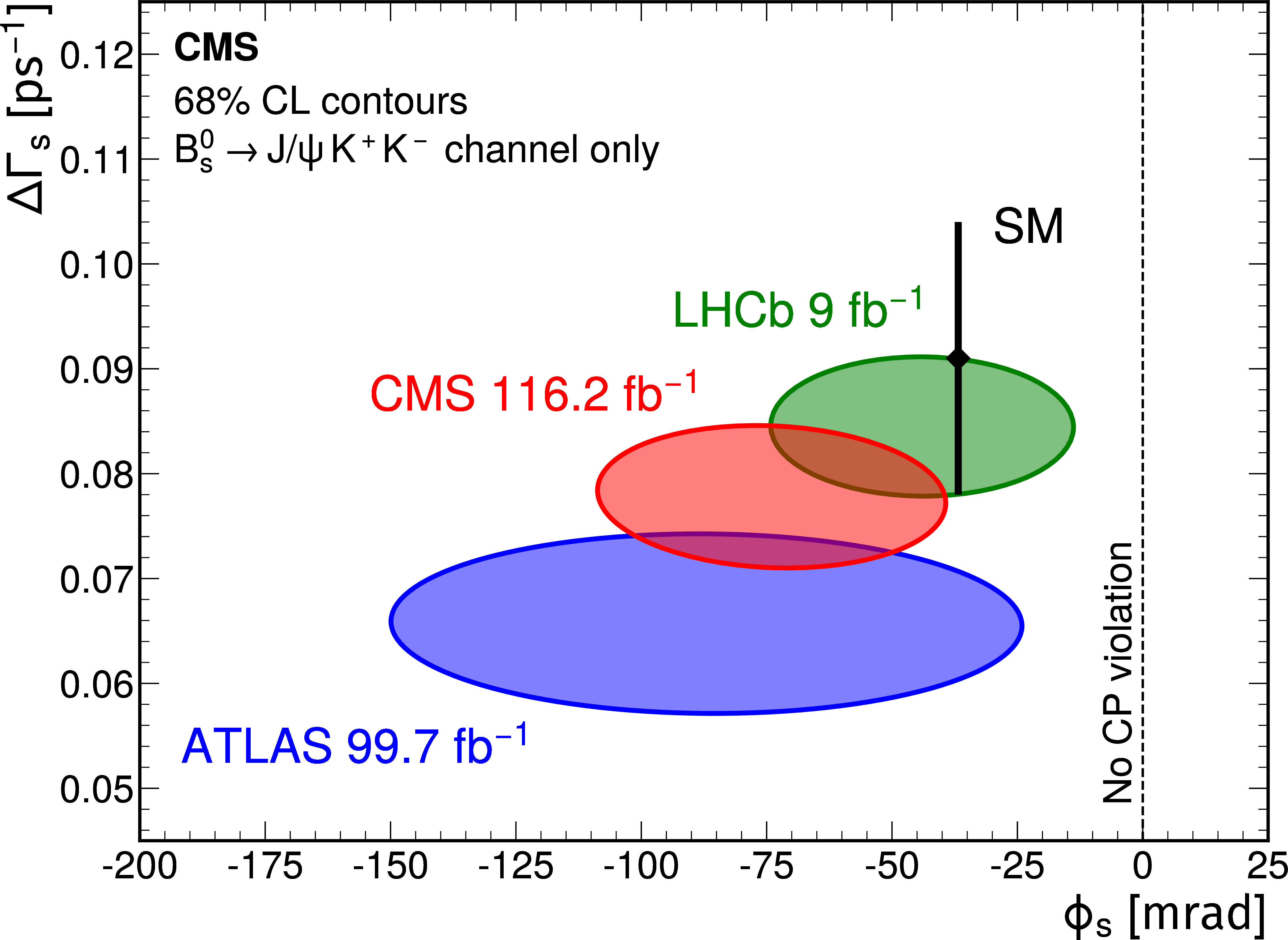}
   \caption{Determination of $\phi_{B_s}$ from  the time evolution of 
the $B_s \to \psi \phi$ decay, as obtained by 
   ATLAS, CMS, and LHCb. Figure from Ref.~\cite{CMS:2024znt}.}
  \label{fig:Phis}
\end{figure}

The golden channel for the measurement of  the CP-violating phase of  
$B_s$--$\bar B_s$ mixing is the time-dependent analysis  
of the $B_s(\bar B_s) \to \psi \phi$ decay. At the quark level 
$B_s \to \psi \phi$  share the same virtues of  $B_d \to \psi K$ 
(partonic amplitude of the type  $b\to c \bar c s$), 
which is used to extract the phase of $B_d$--$\bar B_d$ mixing.
However, there a few points which makes this measurement much 
more challenging:
\begin{itemize}
\item The $B_s$ oscillations are much faster ($\Delta m_{B_s}/\Delta m_{B_d} 
\approx F^2_{B_s}/F^2_{B_d} | V_{ts}/V_{td}|^2$ $\approx 30$), making the time-dependent analysis
quite difficult (and essentially inaccessible at $B$ factories).
\item Contrary to $\ket{\psi K}$, which has a single angular momentum and is a pure CP eigenstate, 
the vector-vector state $\ket{\psi \phi}$ produced by the $B_s$ decay has different angular momenta, 
corresponding to different CP eigenstates. These must be disentangled with a proper angular 
analysis of the final four-body final state $\ket{(\ell^+\ell^-)_{\psi} (K^+K^-)_{\phi}}$.
To avoid contamination from the nearby $\ket{\psi f_0}$ state, the fit should include also
a $\ket{(\ell^+\ell^-)_{\psi} (K^+K^-)_{S-{\rm wave}}}$ component, for a total of ten independent
(and unknown) weak amplitudes.
\item Contrary to the $B_d$ system, the width difference cannot be neglected in the 
$B_s$ case, leading to an additional key parameter to be included in the fit. 
\end{itemize}
Modulo the experimental difficulties listed above, the process is theoretically clean and 
a complete fit of the decay distributions should allow the extraction of 
\be
\lambda^{B_s}_{\psi \phi} \approx - e^{- i \phi_{B_s} }~, 
\ee
where the SM prediction is 
\be 
\phi_{B_s}^{\rm SM}  = - {\rm arg} \frac{ (V_{tb}^* V_{ts})^2 }{ |V_{tb}^* V_{ts}|^2 } \approx  -0.035~.
\ee
In the last few years all the three LHC experiments (ATLAS, CMS and LHCb) have been able to measure $\phi_{B_s}$ from the time evolution of 
the $B_s \to \psi \phi$ decay. The (impressive) comparison of their results with the SM expectation is shown in Fig.~\ref{fig:Phis}.

\subsection{Rare decays: generalities}
\label{sect:rare}

The effective Lagrangians relevant to rare processes 
of the type  $b\to s(d)  + \bar f f$,  where $f = q, \ell, \nu$,
 are slightly more complicated than those discussed so far. 
The conceptual steps necessary to derive their structure 
are the same as those discussed in the case of 
$\Delta F=2$  and semileptonic processes.  
However, the procedure is more lengthy 
(and the operator list is longer) given the appearance a new 
class of diagrams, the so-called {\em penguin} diagrams.\footnote{ {\em John Ellis will tell you the interesting story around this name}\ldots } 
In the following we will briefly review the structure of these effective Lagrangians
without discussing their  derivation in great detail 
(an exhaustive discussion can be fount in Ref.~\cite{Buchalla:1995vs}).

For concreteness, let's give a first look to non-leptonic processes where the underlying partonic 
transition is  $b\to s + \bar q q$. In this case the relevant effective Lagrangian 
can be written as
\be
\cL_{b \to s}^{\rm non-lept} = - 4 \frac{G_F}{\sqrt{2}} \left(
\sum_{q=u,c} \lambda_q^s  \sum_{i=1,2} C_i(\mu) Q^q_i(\mu) 
-\lambda_t^s \sum_{i=3}^{10} C_i(\mu) Q_i(\mu)\right)~,
\label{eq:hdb1}
\ee
where $\lambda^{s}_{q}=V^*_{qb} V_{qs}$, and the operator basis is 
\be
\begin{array}{ll}
Q^q_{1} = {\bar b}_L^\alpha\gamma^\mu q_L^\alpha\, {\bar q}_L^\beta\gamma_\mu s_L^\beta~, &
Q^q_{2} = {\bar b}_L^\alpha\gamma^\mu q_L^\beta\, {\bar q}_L^\beta\gamma_\mu s_L^\alpha~, \\
Q_{3} = {\bar b}_L^\alpha \gamma^\mu s_L^\alpha\, \sum_q {\bar q}_L^\beta\gamma_\mu q_L^\beta~, &
Q_{4} = {\bar b}_L^\alpha \gamma^\mu s_L^\beta\, \sum_q {\bar q}_L^\beta\gamma_\mu q_L^\alpha~, \\
Q_{5} = {\bar b}_L^\alpha \gamma^\mu s_L^\alpha\, \sum_q {\bar q}_R^\beta\gamma_\mu q_R^\beta~,  
\qquad\qquad &
Q_{6} = {\bar b}_L^\alpha \gamma^\mu s_L^\beta\, \sum_q {\bar q}_R^\beta\gamma_\mu q_R^\alpha~, \\
Q_{7} = \frac{3}{2}{\bar b}_L^\alpha \gamma^\mu s_L^\alpha\, \sum_q e_q {\bar q}_R^\beta\gamma_\mu q_R^\beta~, &
Q_{8} = \frac{3}{2}{\bar b}_L^\alpha \gamma^\mu s_L^\beta\, \sum_q e_q {\bar q}_R^\beta\gamma_\mu q_R^\alpha~, \\
Q_{9}^{[q]} = \frac{3}{2}{\bar b}_L^\alpha \gamma^\mu s_L^\alpha\, \sum_q e_q {\bar q}_L^\beta\gamma_\mu q_L^\beta~, &
Q_{10}^{[q]} =  \frac{3}{2}{\bar b}_L^\alpha \gamma^\mu s_L^\beta\, \sum_q e_q {\bar q}_L^\beta\gamma_\mu q_L^\alpha~,\\
\end{array}
\label{eq:basis}
\ee
with $\{\alpha,\beta\}$ and $e_q$ denoting color indexes
the electric charge of the quark $q$, respectively. 

Out of these operators, only $Q^c_1$ and $Q^u_1$ are generated at the tree-level
by the $W$ exchange. Indeed, comparing with the tree-level structure in~(\ref{eq:LFermi0}), 
we find 
\be 
C^{u,c}_1(m_W) = 1 + \cO(\alpha_s,\alpha)~, \qquad 
 C^{u,c}_{2-10}(m_W) = 0 + \cO(\alpha_s,\alpha)~.
\ee
However, after including RGE effects and running down to $\mu\approx m_b$, 
both $C^{u,c}_1$ and $C^{u,c}_2$ become $\cO(1)$ and running further down
to $\mu\approx 1$~GeV also  $C_{3-6}$ become $\cO(1)$.

\begin{figure}[t]
\begin{center}
\includegraphics[width=4.6cm]{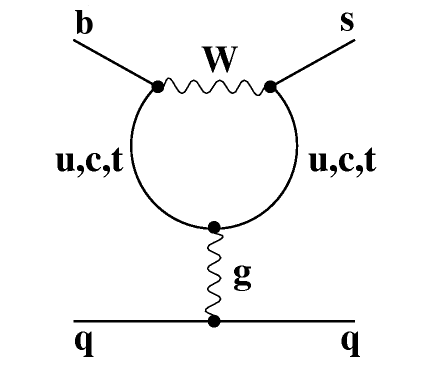}
\hskip 1 cm 
\includegraphics[width=3.5cm]{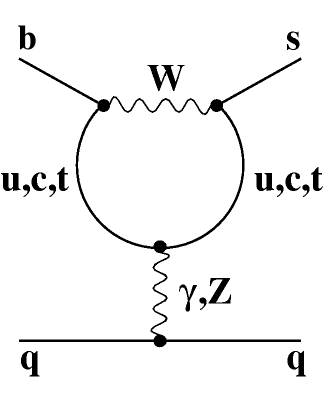}
\end{center}
\caption{One-loop penguins diagrams: QCD penguins (left) and EW penguins (right).}
\label{fig:penguins}
\end{figure}

The operators $Q_{3-8}$ and $Q^{[q]}_{9,10}$ are present  only in  processes of the type 
$b\to s + \bar q q$ or, with a suitable change of quark flavors, in
processes of the type $b\to d + \bar q q$ and $s \to d + \bar q q$.
Only in such processes one-loop topologies 
of the type in Fig.~\ref{fig:penguins} are allowed.
These  operators are not present in processes 
such as   $b\to c \bar \ell \bar \nu$ or $b\to c \bar u d$, which involve four different 
fermion species. 

The operators $Q_{3-6}$ are generated at the one-loop 
level by the QCD penguins (Fig.~\ref{fig:penguins} left),
while  $Q_{7,8}$ and $Q^{[q]}_{9,10}$ are generated by the electroweak (EW) penguin diagrams 
of the type in Fig.~\ref{fig:penguins} (right) and by related box diagrams
(similar to those appearing in $\Delta F=2$ amplitudes).
These diagrams involve all three types of up-type quarks inside the 
loops. However, since  $\sum_{q=u,c,t} V^*_{qb} V_{qs} =0$, we can 
always eliminate one CKM combination, as we have already seen in the case of the 
$\Delta F=2$ amplitudes.  Moreover, if the amplitude is
regular in the limit of vanishing up-quark mass, the mass-independent
part of the loop amplitude cancels --once more as a manifestation of the 
GIM mechanism~\cite{Glashow:1970gm} or the dominance of the top Yukawa coupling-- leaving a sizeable contribution 
only in the part of the amplitude proportional to  $V^*_{tb} V_{ts}$:
\bea
\cA_{\rm peng} &=& 
 \sum_{q=u,c,t} V^*_{qb} V_{qs}~ \cA\left(\frac{m^2_q}{m^2_W}\right) \no \\
&=& V^*_{cb} V_{cs} \left[ \cA\left(\frac{m^2_c}{m^2_W}\right) - \cA\left(\frac{m^2_u}{m^2_W}\right) \right] + V^*_{tb} V_{ts} \left[ \cA\left(\frac{m^2_t}{m^2_W}\right) - \cA\left(\frac{m^2_u}{m^2_W}\right) \right] \no\\
&\approx &   V^*_{tb} V_{ts}~ \cF\left(\frac{m_t}{m_W}\right)~.
\eea
This is why in Eq.~(\ref{eq:hdb1}) the coefficients $C_{3-10}$
are multiplied only by the CKM combination  $\lambda^{s}_{t}$.

The coefficients of the penguin operators at the electroweak 
scale are potentially more sensitive to NP with respect 
to the initial values of the other four-quark operators
(being free from tree-level SM contributions). However, it is 
hard to distinguish their contribution from those of the other 
four-quark operators in non-leptonic processes. Moreover, 
the relative contribution from long-distance 
physics (running down from  $m_W$ to $m_b$) 
is sizable and can dilute the interesting short-distance 
information at low energies.

The situation is quite different for the so-called flavor-changing neutral-current  (FCNC) processes,
namely $b\to s \ell^+ \ell^-$,  $b\to s \bar \nu \nu$,  and $b\to s \gamma $ (and similarly for $b\to d$ and $s\to d$).
Here all the  four-quark operators have vanishing tree-level matrix elements. In all cases where the decay amplitude 
is dominated by the contributions of the electroweak penguins the process provides an information on 
physics occurring at high scales with no tree-level SM contribution, hence it provides  an interesting window on physics  beyond the SM.

\subsection{Effective Lagrangian for $\bsll$ decays}
 
For $b\to s$ transitions with a photon or a charge-lepton pair in the final state additional
dimension-six operators must be included in the basis, 
\be
\cL_{b \to s}^{\rm rare} = \cL_{b \to s}^{\rm non-lept} +  4 \frac{G_F}{\sqrt{2}} \lambda_t^s
\left(C_{7}^\gamma Q_{7}^{\gamma} + C_{8}^{g} Q_{8}^{g} +  C^\ell_{9} Q^\ell_{9} + 
 C^\ell_{10} Q^\ell_{10}\right)~,
\ee
where 
\begin{eqnarray}
&& Q_{7}^\gamma = \frac{e}{16\pi^2} m_b {\bar b}_R^\alpha\sigma^{\mu\nu} F_{\mu\nu} s_L^\alpha~, \qquad\
Q_{8}^g = \frac{g_s}{16\pi^2} m_b {\bar b}_R^\alpha\sigma^{\mu\nu} G_{\mu\nu}^A T^A s_L^\alpha~,   \nonumber\\
&& Q^\ell_{9} = \frac{1}{2}{\bar b}_L^\alpha \gamma^\mu s_L^\alpha\, \bar \ell \gamma_\mu \ell~, \qquad\qquad\quad 
Q^\ell_{10} = \frac{1}{2}{\bar b}_L^\alpha \gamma^\mu s_L^\alpha\, \bar \ell \gamma_\mu\gamma_5 \ell~,
\label{eq:radbasis}
\end{eqnarray}
and  $G^A_{\mu\nu}$ ($F_{\mu\nu}$) is the gluon (photon) field strength tensor. 
The initial conditions of these 
operators are particularly sensitive to NP and, contrary to 
non-leptonic processes, in this case is easier to isolate their 
contribution in low-energy observables.
The ``cleanest'' case is $C_{10}$, which do not mix with any of the 
four-quark operators listed above and is not renormalised by 
QCD corrections:
\be 
\left. C^\ell_{10}(m_W) \right|^{\rm SM} = \frac{g^2}{8\pi^2} \frac{x_t}{8}\left[ \frac{4- x_t}{1-x_t}
  +\frac{3x_t}{(1-x_t)^2}\;\ln x_t\right]~, \qquad x_t = \frac{m_t^2}{m_W^2}~.
\label{eq:Y0}
\ee  
NP effects at the TeV scale could easily modify this result, and this deviation 
would directly show up in low-energy observables sensitive to $C^\ell_{10}$, such as the 
branching ratio of the rare decays  $B_{s,d} \to \ell^+\ell^-$.

The operator basis in  Eqs.~(\ref{eq:basis}) and (\ref{eq:radbasis}) is a complete basis to describe rare FCNC decays within the SM.
However, when going beyond the SM new type of effects can be generated. In particular,  right-handed flavor-changing 
quark currents could appear. These would be described by operators identical to those in Eqs.~(\ref{eq:basis}) and (\ref{eq:radbasis}) 
with the exchange $b_L \leftrightarrow  b_R$  and $s_L \leftrightarrow  s_R$. These are usually denoted $Q^\prime_i$ operators.
Within the SM the  $C^\prime_i$ are not completely negligible, but are proportional to light-quark Yukawa couplings hence can be safely
neglected.
 
A further important prediction of the SM is that the Wilson coefficients of  $Q^\ell_9$ and $Q^\ell_{10}$ are lepton-flavor universal.
Indeed the lepton label of these operators is usually omitted in SM analyses.
This property goes under the name of Lepton Flavor Universality (LFU). Similarly to the case of the operators with right-handed currents,
the SM does not predict exact LFU, but LFU violating effects of SM origin are proportional to the lepton Yukawa couplings hence 
can be safely neglected, especially for $\ell=e,\mu$.

%In order to analyze NP effects that violate LFU, it is convenient to distinguish LFU-breaking contributions from universal NP corrections to the Wilson coefficients. We choose   to define the universal corrections using the Wilson coefficients of the electron modes as reference, i.e.\
%\be
%\Delta \cC_{i}^U \equiv \cC^e_i -\cC^{\rm SM}_i  \qquad [i=9,10] \,, 
%\ee
%such that the LFU-breaking terms can be defined as
%\be
%\Delta \cC_i^{\mu} \equiv \cC_i^\mu - \cC_i^e  =
 %\cC^\mu_i - ( \cC^{\rm SM}_i +\Delta \cC_{i}^U )  \qquad [i=9,10] \,.
 %\label{eq:DCidef}
%\ee

\subsection{Observables in exclusive $\bsll$ decays}

\begin{figure}[t]
\centering
\vskip -4 true cm
\includegraphics[width=0.9\textwidth]{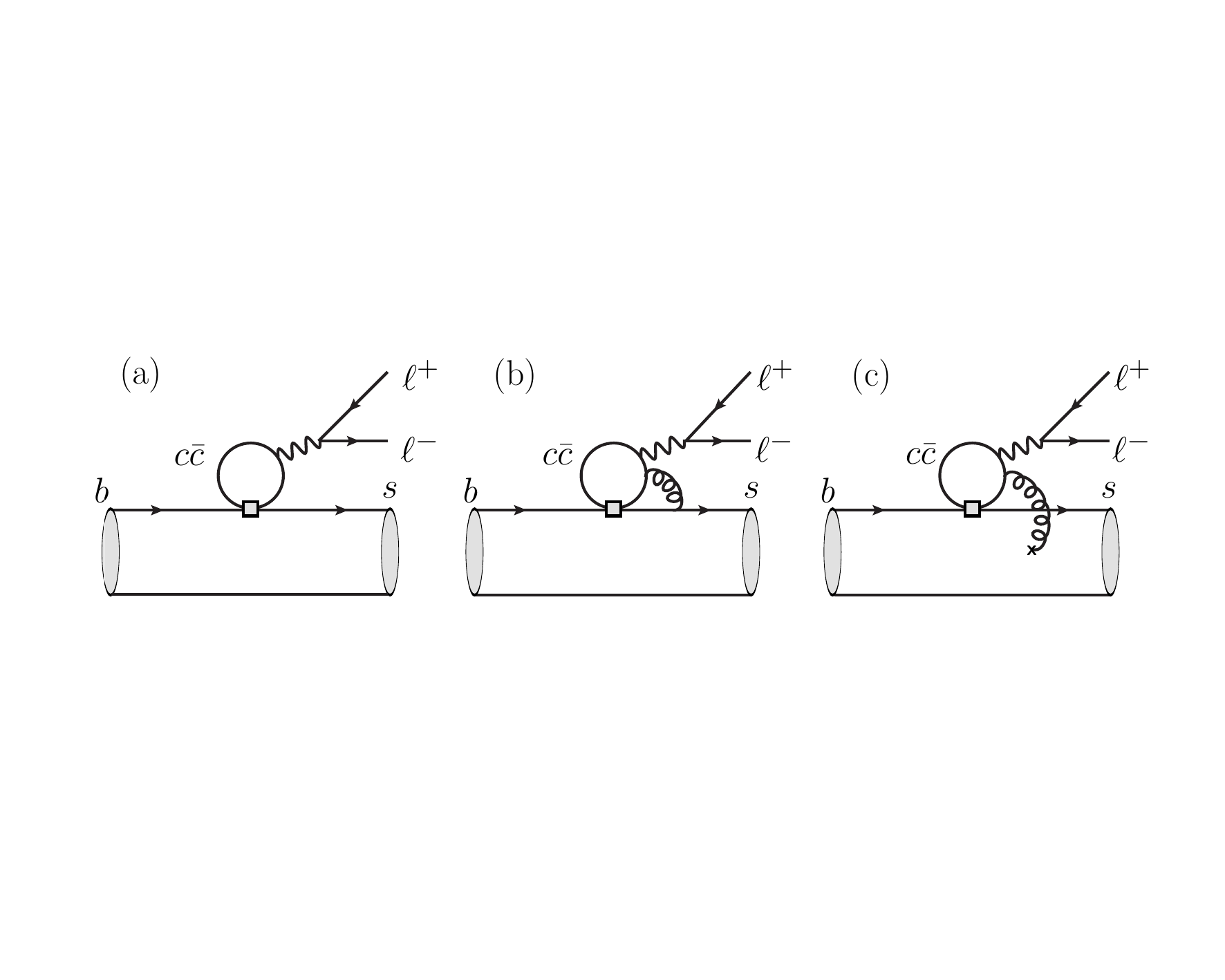} 
\vskip -4 true cm
\caption{  \label{fig:cctopologies}
Representative diagrams for the matrix element of the leading four-quark operators in $B \to X \ell^+\ell^-$ decays. }
\end{figure}

Following the general discussion in the first lecture (section~\ref{sect:Heff}),  after having identified the $\bsll$  
effective Lagrangian, in order to compute the decay amplitudes for a specific exclusive hadronic process, 
such as  $B_s  \to \ell^+\ell^-$ or $B^+ \to K^+ \ell^+\ell^-$, 
we need to evaluate the hadronic matrix elements of the effective operators. 
For processes with at most one hadron in the final state, 
and considering only operators bilinear in the quark fields this is  quite simple,
(at least if we neglect higher-order terms in the electroweak couplings).  
In this limit we can indeed  
factorise the complete matrix element into the hadronic matrix element of a quark current, 
and a leptonic term. For instance, in the case of the two leading FCNC 
operators $Q^\ell_9$ and $Q^\ell_{10}$ we can write 
\be
\langle \ell^+ \ell^-  X  \vert  Q^\ell_{9,10}  \vert  B  \rangle  =   \langle X \vert\, \bar s_L  \gamma^\mu  b_L\,  \vert  B \rangle\, 
\times \left\{  \ba{ll}  \bar \ell \gamma_\mu \ell &\quad [Q^\ell_9]\,, \\ \bar \ell \gamma_\mu\gamma_5 \ell
&\quad [Q^\ell_{10}]\,. \ea \right.
\ee
This factorization holds to all orders in QCD. It is violated by structure-dependent QED corrections, 
but in this case the effect is usually very small.

In order to evaluate the hadronic part, namely $\langle X \vert\, \bar s_L  \gamma^\mu  b_L\,  \vert  B \rangle$,  we need appropriate form factors.
If $| X \rangle$ is not the vacuum, these  need to be evaluated over the whole kinematical regime of the dilepton invariant 
mass $q^2=(p_{\ell^+}+p_{\ell^-})^2$. 
For $B \to K$ and $B \to K^*$ decays these have been computed both using 
light-cone sum rules (LCSR) and lattice QCD. 
Further precision can be gained by performing combined fits of the lattice results valid at high $q^2$
and LCSR results valid at low $q^2$ (see e.g.~Ref.~\cite{Bharucha:2015bzk,Horgan:2015vla,Capdevila:2017ert,Gubernari:2018wyi,Parrott:2022rgu,Monceaux:2023byy,Gubernari:2023puw}).

\begin{figure}[t]
\begin{center}
\includegraphics[width=0.8\textwidth]{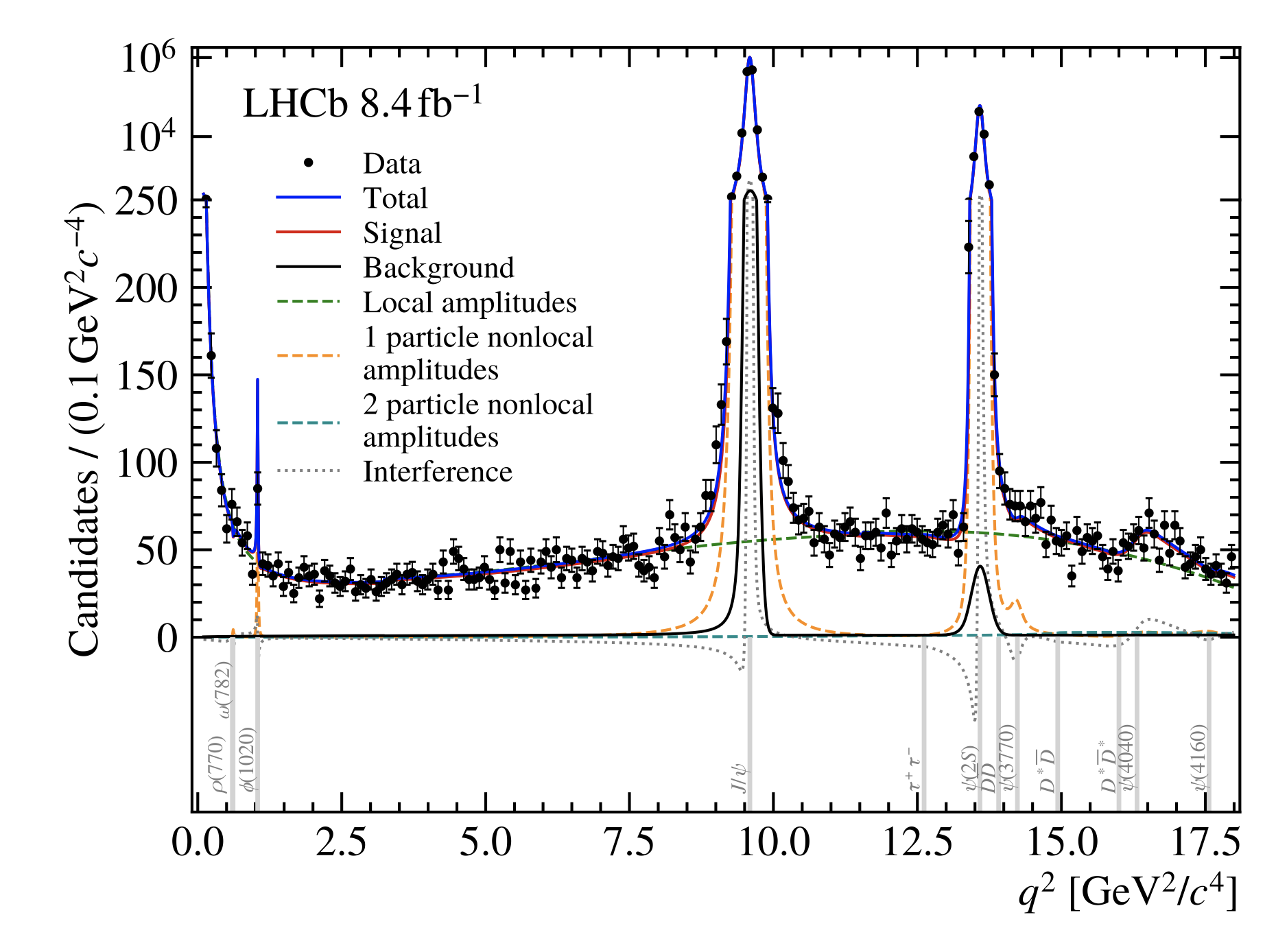}
\end{center}
\vskip - 1cm
\caption{Dilepton spectrum in $B \to K^* \mu^+\mu^-$ as measured by LHCb~\cite{LHCb:2024onj}.
 \label{fig:LHCb-BR-hmumu} }
\end{figure}

Beyond the form factors, an important source of theoretical uncertainties are the matrix elements  
of four-quark operators,  and in particular those of the leading operators $Q^c_{1,2}$ which, 
being generated at the tree level,  have large numerical coefficients. 
The structure of representative hadronic matrix element for these operators are illustrated in Fig.~\ref{fig:cctopologies}.
For  $q^2$ around the masses of the narrow $c\bar c$ resonances, these diagrams give rise to large 
long-distance effects (associates to the on-shell production of the resonances)
which cannot be estimated in perturbation theory. This is evident by looking at the dilepton invariant mass spectrum 
measured by LHCb in the  $B \to K^* \mu^+\mu^-$ mode, shown in Fig.~\ref{fig:LHCb-BR-hmumu}. This is why the region
$8~{\rm GeV}^2 \leq q^2 \leq 15~{\rm GeV}^2$ there is no chance to preform precise tests of 
short-distance dynamics.

Far from the narrow resonance regions, the rate is dominated by the matrix elements of $Q^\ell_{9,10}$ and $Q^\gamma_7$,
as can be seen looking at the dark-green dashed line in Fig.~ \ref{fig:LHCb-BR-hmumu} (denoted ``Local amplitudes").
However, a reliable estimate of the matrix elements in Fig.~\ref{fig:cctopologies} is needed to perform precise tests of short-distance dynamics. 
The diagrams of the types (a) and (b) denote the factorizable part of the matrix element that,
far from the resonance region, can be estimated reliably in perturbative QCD and dispersive methods.
The largest concern are the non-factorizable (and non-local) corrections of the type (c).
These effects have been estimated in~\cite{Khodjamirian:2010vf,Khodjamirian:2012rm} 
in the low-$q^2$ region via a power expansion in $\Lambda_{\rm QCD}/m_c$
and turn out to be quite small, in both $B\to K$ and $B\to K^*$ decays. 
A more general approach based on analyticity, data, and perturbative 
constraints~\cite{Bobeth:2017vxj,Gubernari:2020eft,Gubernari:2022hxn,Gubernari:2023puw},
as well as an explicit estimate of long-distance corrections in terms of meson states~\cite{Isidori:2024lng},
further confirm the smallness of these non-local matrix elements. 
A data-driven analysis aimed at isolating possible non-factorizable long-distance corrections in the
 data, by looking at their possible dependence from $q^2$ and/or from the specific final state, is also 
 compatible with the hypothesis that these effects are not large~\cite{Bordone:2024hui} (at most few $\%$ relative to the leading 
 local amplitude).  
However, the uncertainty due to these non-perturbative contributions
remains a source of concern due to the lack of a systematic control of all the approximations 
entering their evaluation~\cite{Ciuchini:2020gvn,Ciuchini:2021smi,Hurth:2023jwr}.

\begin{figure}[t]
\begin{center}
\includegraphics[width=0.7\textwidth]{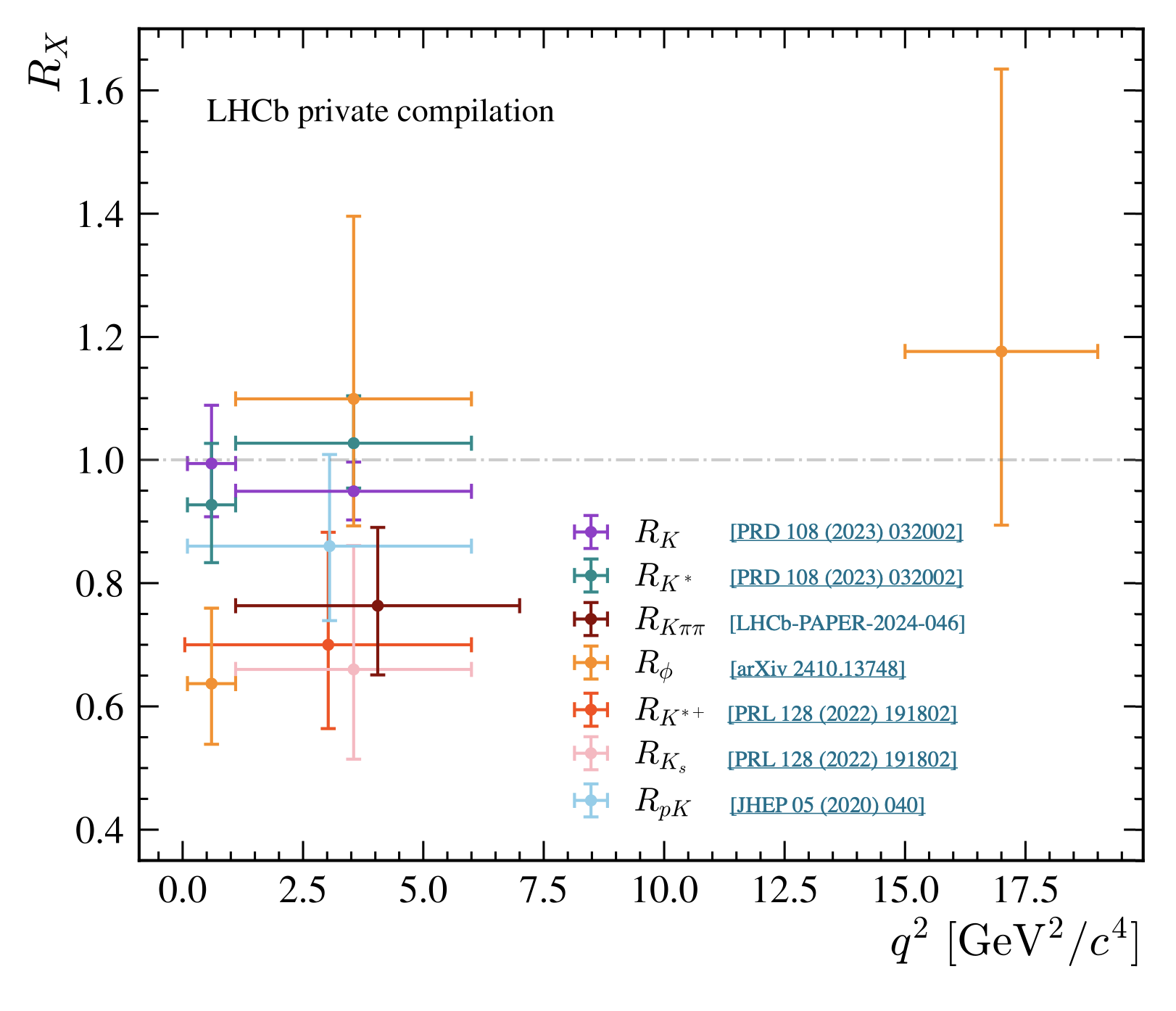}
\end{center}
\vskip - 0.5 cm
\caption{ Compilation of the latest measurements of $\mu/e$ LFU ratios obtained by the LHCb Collaboration~\cite{LHCb:2019efc,LHCb:2021lvy,LHCb:2022vje,LHCb:2024rto,LHCb:2024yci}. Figure from Ref.~\cite{SCelani}.
 \label{fig:RX} }
\end{figure}

An important observation is that the matrix elements of four-quark operators cannot induce violations of LFU and cannot induce 
amplitudes where the leptonic part of the matrix element is an axial current.
This allows us to define the set of so-called {\em  clean observables}, i.e.~observables which are 
insensitive to the sizeable theoretical uncertainties associated to $c\bar c$ re-scattering 
at least up to higher-order QED corrections. These consist in particular of two classes: the  $\mu/e$ LFU  ratios  and the 
purely leptonic $B_{s,d} \to \ell^+\ell^-$ rates.

\begin{itemize}
\item{\underline{The  $\mu/e$ LFU  ratios.}} These are defined as
\begin{equation}
    R_{X} \equiv  \dfrac{\displaystyle\int_{q^2_\mathrm{min}}^{q^2_{\rm max}} \dfrac{{\rm d} \Gamma ( B \to X \mu^+\mu^-)}{ {\rm d} q^2} {\rm d} q^2}{\displaystyle\int_{q^2_{\rm min}}^{q^2_{\rm max}} \dfrac{ {\rm d} \Gamma (B \to X e^+e^-)}{ {\rm d} q^2} {\rm q^2}}~.
    \label{eq:RX}
\end{equation}
Neglecting phase-space effects, which are negligible for $q^2_{\rm min}\geq 1$~GeV, and QED corrections, the SM expectation is
$R_{X}^{\rm SM}=1$, while extensions of the SM can leads to sizeable deviations~\cite{Hiller:2003js}. Within the SM,  the only source of theoretical uncertainty are QED corrections. These are completely negligible if the measurements are fully inclusive with respect to the electromagnetic radiation~\cite{Bordone:2016gaq,Isidori:2020acz}. 
This is not the case for the measurements performed by LHCb; however, the effect is corrected for by the experiment 
assuming a pure bremsstrahlung spectrum. 
By doing so, the residual error due to QED corrections  does not exceed  $1\%$ for $q^2_{\rm min}\geq 1$~GeV~\cite{Bordone:2016gaq,Isidori:2020acz}. 
 
\begin{figure}[t]
\begin{center}
\includegraphics[width=1.0\textwidth,height=0.33\textwidth]{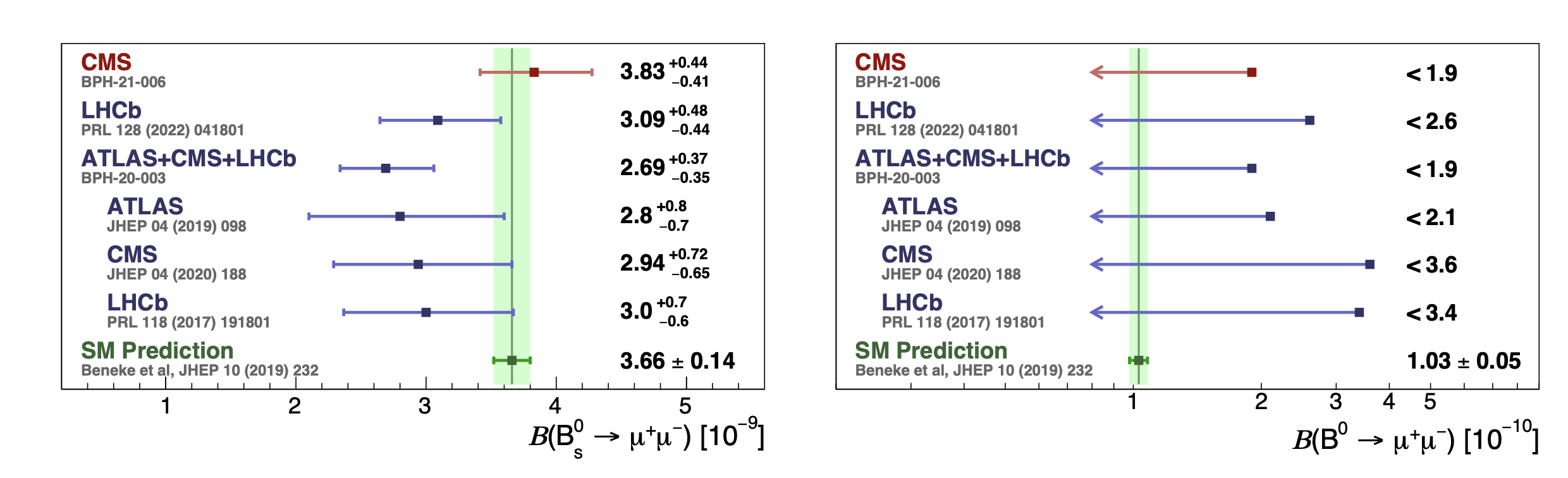}
\end{center}
\vskip  -0.5 cm
\caption{Compilation of the latest measurements of $\cB(B_s \to \mu^+\mu^-)$ and $\cB(B_d \to \mu^+\mu^-)$ by the LHC experiments. Figure from Ref.~\cite{Simone:2024vzs}.
 \label{fig:Bsmm} }
\end{figure}

\item{\underline{$\cB(B_{s,d} \to \ell^+\ell^-)$.}}
The conservation of angular momentum implies that the two leptons in these processes must have opposite helicity.
This, in turn, implies the decay amplitude is helicity suppressed and does not receive tree-level contributions by
vector-current operators, such as $Q_9$, or single-photon exchange amplitudes, such as those in Fig.~\ref{fig:cctopologies}.
Within the SM these processes are completely dominated by the contribution of $Q_{10}$
and can be estimated with high accuracy~\cite{Buchalla:1995vs,Buras:2012ru,Bobeth:2013uxa}
in terms of the meson decay constants ($f_{B_{d,s}}$), defined by
\be
\langle 0 \vert\, \bar q  \gamma^\mu \gamma_5  b \,  \vert  B_q (p) \rangle\, =  i f_{B_q} p^\mu~. 
\ee
A systematic analysis including higher-order QED-QCD corrections has been presented in~\cite{Beneke:2017vpq,Beneke:2019slt},
where the overall relative theory uncertainty on $\cB(B_s \to \mu^+\mu^-)_{\rm SM}$  is estimated to be $4\%$.
This is due to a combination of parametric uncertainties,  from  $f_{B_s}$ and CKM matrix elements
(leading contribution), and the higher-order QED-QCD corrections.
\end{itemize}
Until 2021, there was significant tension between SM predictions and observations in both sets of clean observables, most notably in $R_K$, 
$R_{K^*}$, and   $\cB(B_s \to \mu^+\mu^-)$. This fact generated a great interest from the model-building perspective, stimulating the focus on theories with violations of lepton universality. From the theoretical point of view, the interest in this class of models is still very high, as we will discuss in the next lecture. 
On the the other hand, the situation has changed significantly from the phenomenological perspective: as shown in Fig.~\ref{fig:RX}--\ref{fig:Bsmm}, nowadays the tension almost disappeared in both cases. It is worth stressing that  this does not mean that deviations from the SM cannot be detected in the future through these observables: in both cases the current experimental errors are well above the level of the irreducible theory uncertainties. Deviations from the SM at the $5\%$--$10\%$ level are still possible in both cases, and actually are welcome in motivated extensions of the SM. If present, these would be within the reach of future measurements.

%\newpage 
\section{Lecture III:  Flavor physics beyond the SM}

\subsection{The SM as an effective theory and the flavor problem}

Despite the impressive phenomenological success of the SM in flavor and electroweak physics, 
as well as the absence of signals of new particles in direct searches, there are various 
convincing arguments which motivate us to consider this model only as the low-energy limit of a more complete theory. 
These include the evidence of neutrino masses and dark matter, as well as more theoretical arguments such as the instability under quantum corrections of the Higgs sector (the so-called electroweak  or Higgs {\em hierarchy problem}) and the lack of a consistent inclusion of gravity at the quantum level.

Assuming that the new degrees of freedom are heavier than the SM ones,
we can integrate them out and describe physics beyond the SM in full
generality by means of an effective theory approach.
As we have seen in the previous lectures,
integrating out heavy degrees of freedom is a procedure
often adopted also within the SM: it allows us to 
simplify the evaluation of amplitudes which involve 
different energy scales. This approach is indeed
a generalization of the Fermi theory of weak interactions,
where the dimension-six four-fermion operators describing 
weak decays are the results of having integrated out 
the $W$ field. The only difference when applying this 
procedure to physics beyond the SM is that in this case,
as also in the original work by Fermi, we don't know the 
nature of the degrees of freedom we are integrating out. 
This imply  we are not able to determine a priori the 
values of the effective couplings of the higher-dimensional 
operators. The advantage of this approach is that it 
allows us to analyse all realistic extensions of the SM 
in terms of a limited number of parameters 
(the coefficients of the higher-dimensional operators). 
The drawback is the impossibility to establish correlations of NP effects at low and high energies.

The Lagrangian of the SM considered as an effective theory, often denoted SMEFT (see Ref.~\cite{Brivio:2017vri,Isidori:2023pyp} for a review), 
can be decomposed as follows
\be
\cL_{\rm SMEFT} = \cL^{\rm SM}_{\rm gauge} + \cL^{\rm SM}_{\rm Higgs} + 
\cL^{\rm SM}_{\rm Yukawa}
+\Delta \cL_{d > 4}~,
\ee
where $\Delta \cL_{d > 4}$ denotes the series of higher-dimensional 
operators, of canonical dimension $d$, invariant under the SM gauge group:
\be
\Delta \cL_{d > 4}  = ~\sum_{d > 4} ~\sum_{n=1}^{N_d} ~
\frac{c^{(d)}_n}{\Lambda^{d-4}} \cO^{(d)}_n ({\rm SM~fields}). 
\label{eq:DL_eff}
\ee
Interestingly enough, there is a single operator with dimension $d=5$ invariant under the 
SM gauge group: it is the Weinberg operator describing non-vanishing neutrino masses.
On the other hand, a long list of operators appears at  $d=6$, and for them there is 
no positive experimental  evidence yet. These  are the operators we  ``hunt for" in flavor physics.

If NP appears at the TeV scale, as we expect from the stabilization 
of the Higgs sector, the scale 
$\Lambda$ cannot exceed a few TeV.  Moreover, if the underlying 
theory is natural  (no fine-tuning in the coupling constants), we 
expect $c^{(d)}_i=O(1)$ for all the operators which 
are not forbidden (or suppressed) by symmetry arguments. 
The observation that this expectation is not  
fulfilled by several dimension-six operators contributing 
to flavor-changing processes is often denoted 
as the NP {\em flavor problem}.

\subsection{Bounds on the scale of New Physics from $\Delta F=2$
processes}
\label{sect:DF2bounds}

The best way to quantify the NP flavor problem is obtained 
by looking at consistency of the tree- and loop-mediated 
constraints on the CKM matrix 
discussed in the first part of these lectures.

In first approximation we can assume that NP effects are
negligible in processes which are dominated by tree-level 
amplitudes. Following this assumption, the values of 
$|V_{us}|$, $|V_{cb}|$, and $|V_{ub}|$, as well as the 
constraints on $\alpha$ and $\gamma$ 
are essentially NP free. As can be seen in Fig.~\ref{fig:UT},
this implies we can determine completely the CKM matrix assuming generic 
NP effects in loop-mediated amplitudes.
We can then use the measurements of observables which are
loop-mediated within the SM to bound the couplings of 
effective NP operators in Eq.~(\ref{eq:DL_eff})
which contribute to these observables at the tree level.

The loop-mediated constraints shown in Fig.~\ref{fig:UT}
are those from the mixing of $B_d$, $B_s$, and $K^0$
with the corresponding anti-particles
(generically denoted as $\Delta F=2$ amplitudes).
As we have seen in the first lecture, within the SM 
these amplitudes are generated by 
box amplitudes dominated by the top-quark exchange,
leading to 
\be
 \cM_{\Delta F=2}^{\rm SM} \approx  \frac{ G_F^2 m_t^2 }{16 \pi^2} ~ V_{3i}^* V_{3j} ~
    \langle \bar  M |  (\bar d_L^i \gamma^\mu d_L^j )^2  
| M \rangle \times  F\left(\frac{m_t^2}{m_W^2}\right)
\qquad [M = K^0, B_d, B_s]~,
\ee
where $F$ is a loop function of order one 
($i,j$ denote the flavor indexes of the meson
valence quarks).

Magnitude and phase of all these
mixing amplitudes have been determined 
with good accuracy from experiments, showing 
good consistency with the SM predictions.
We can thus deduce that  the magnitude of the new-physics amplitude
cannot exceed, in size, the SM contribution. 
To translate this information into bounds on the scale 
of new physics, let's consider the following set of 
$\Delta F=2$ dimensions-six operators
\be
\cO_{\Delta F=2}^{ij} = (\bar Q_L^i \gamma^\mu Q_L^j )^2 ~,
\qquad Q^i_L =\left(\ba{c} u^i_L \\ d^i_L \ea \right)~,
\label{eq:dfops}
\ee
where $i,j$ are flavor indexes in the 
mass-eigenstate basis of down-type quarks 
(see lecture n.~1).
 
\begin{figure}[t]
\centering
\includegraphics[width=0.65\textwidth]{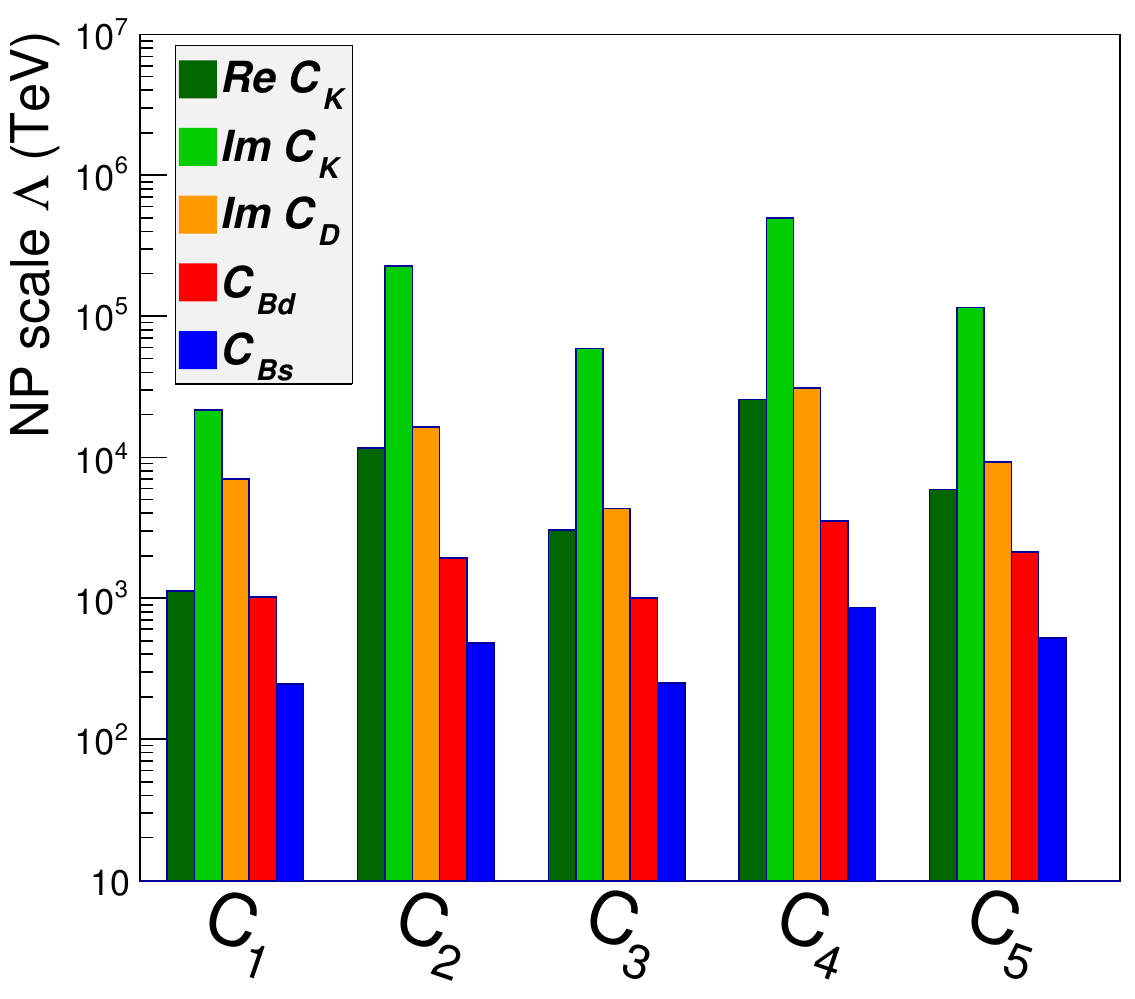}
\caption{Bounds on the scale of different four-fermion operators contributing to 
$|\Delta F|=2$ bounds,  assuming an effective coupling $1/\Lambda^2$~\cite{Alpigiani:2017lpj}.
Here $C_1$ denotes the coefficient of the left-handed operator in Eq.~(\ref{eq:dfops});
see Ref.~\cite{UTfit:2007eik} for  the definition of the other operators. 
The different colors denote different flavor combinations, hence 
different type of $\Delta F=2$  transitions.
  \label{fig:DF22bounds}}
\end{figure}

These operators contribute at the tree-level to 
the meson-antimeson mixing amplitudes.
Denoting $c_{ij}$ the couplings of the non-standard 
operators in (\ref{eq:dfops}), the condition 
$| \cM_{\Delta F=2}^{\rm NP}| <  | \cM_{\Delta F=2}^{\rm SM} |$
implies
\bea
\Lambda < \frac{ 3.4~{\rm TeV} }{| V_{3i}^* V_{3j}|/|c_{ij}|^{1/2}  }
<  \left\{ \ba{l}  
9\times 10^3~{\rm TeV} \times |c_{21}|^{1/2} \qquad {\rm from} \quad 
K^0-\bar K^0
 \\
4\times 10^2~{\rm TeV} \times |c_{31}|^{1/2} \qquad {\rm from}  \quad
B_d-\bar B_d
 \\
7\times 10^1~{\rm TeV} \times |c_{32}|^{1/2} \qquad {\rm from}  \quad
B_s-\bar B_s \ea
\right. 
\label{eq:boundsDF2}
\eea 
A more detailed analysis, separating bounds  for the real and the imaginary parts of the various amplitudes,  
considering also operators  with different Dirac structure, leads to the bounds shown in Fig.~\ref{fig:DF22bounds}.
The main message of these bounds is that NP models with a generic flavor 
structure ($c_{ij}$ of order 1) at the TeV scale are ruled out. If we want 
to keep $\Lambda$ in the TeV range, physics beyond the SM
must have a highly non-generic flavor structure.

\subsection{The role of accidental symmetries}
\label{sect:SymmA}
A key concept in any EFT is that of {\em accidental symmetries},
i.e., symmetries that arise in the lowest-dimensional operators as
indirect consequences of the field content and the symmetries explicitly imposed on the theory.
Within the SMEFT, two well-known examples are baryon number~($B$) and lepton number~($L$).
These are exact accidental global symmetries of the $d=4$ part of the Lagrangian, or the SM:
they do not need to be imposed in the SM because gauge invariance forbids to write any 
$d=4$~operator violating $B$~or~$L$. 

If the accidental symmetries are not respected by the underlying UV~completion,
we expect them to be violated by the higher-dimensional operators. 
The strong bounds on $B$-violating terms from proton stability, and the tiny values of neutrino masses, indicate that such symmetries 
remain almost unbroken in the SMEFT. This observation can be interpreted in a natural way assuming that the fundamental 
interactions responsible for $B$ and $L$ violation appear at very high energy scales, therefore, assuming 
a very high cutoff scale for these operators. 
This is not in contradiction with the possibility of having a lower cutoff scale 
for the $d=6$~SMEFT operators preserving $B$~and~$L$, since the symmetry-preserving sector cannot 
induce violations of the global symmetries. In other words, accidental global symmetries allow us to 
define a stable partition of the tower of effective operators into different sectors characterized 
by different cutoff scales, reflecting a possible multi-scale structure of the underlying theory.
The key point is that this partition is stable with respect to quantum corrections. 

Besides $B$~and~$L$, the SM Lagrangian (or better the SMEFT at~$d=4$) has two additional exact accidental
$U(1)$ global symmetries related to the individual lepton flavors. As we have seen in Eq.~(\ref{eq:YbreaK}),
we can conventionally choose them to be $U(1)_{L_e - L_\mu}$ and $ U(1)_{L_\mu  - L_\tau}$.
 However, a much larger number of {\em approximate accidental symmetries} 
appears in the limit where we neglect the tiny Yukawa couplings of the light families and the small
off-diagonal entries of the CKM  matrix. These approximate flavor symmetries 
are responsible for the smallness of FCNC and $\Delta F=2$ processes.
Despite the precision and the energy scales involved are very different, the situation is 
similar to that of $B$~and~$L$: the experimental bounds on FCNC processes 
imply high cutoff scales for the $d=6$~operators violating the approximate SM flavor symmetries. 

Similarly to the case of exact accidental symmetries, also the approximate accidental symmetries allow us to conceive an underlying multi-scale structure,
separating the symmetry-preserving and symmetry-breaking sectors of the theory
(the maximal scale separation being limited by the size of the explicit symmetry-breaking terms). 
This implies that  the scale of the symmetry-preserving sector of the SMEFT can be as low as few TeV, if at that  
scale not only $B$~and~$L$, but also the tightly constrained accidental flavor symmetries, remain valid,  or are broken only by small symmetry-breaking terms. 
The technical implementation of the concept of small symmetry-breaking terms, in presence of approximate 
(or exact) symmetries in the low-energy sector of the EFT is what we will discuss next, starting from the Minimal Flavor Violation hypothesis.

\subsection{The Minimal Flavor Violation hypothesis}

 The strongest assumption we can make to suppress flavor-changing effects in the  SMEFT is the so-called Minimal Flavor Violation (MFV) 
hypothesis, namely the assumption that  flavor-violating interactions are  generated only by the known structure of the SM Yukawa couplings also beyond the SM.

The MFV hypothesis consists of two ingredients~\cite{DAmbrosio:2002vsn}: 
(i)~a {\em flavor symmetry} and (ii)~a set of {\em symmetry-breaking 
terms}.  The symmetry is noting but the large global 
symmetry ${\mathcal G}_{\rm flavor}$
of the SM Lagrangian in absence of Yukawa couplings
discussed in the lecture n.~1, namely:
\be 
{\mathcal G}_{\rm flavor} = U(3)^5 = U(1)^5 \times  
{\mathcal G}_{q} \times {\mathcal G}_{\ell}~, 
\ee
where 
\be
{\mathcal G}_{q} = {SU}(3)_{Q_L}\times {SU}(3)_{U_R} \times {SU}(3)_{D_R}, \qquad 
{\mathcal G}_{\ell} =  {SU}(3)_{L_L} \otimes {SU}(3)_{E_R}~.
\ee
Since this global symmetry, and particularly 
the ${SU}(3)$ subgroups controlling quark flavor-changing 
transitions, is already broken within the SM, we cannot promote 
it to be an exact symmetry of the NP model. Some breaking would
appear  at the quantum level because of the SM Yukawa interactions.
The most restrictive assumption we can make to ``protect'' in a consistent 
way quark-flavor mixing 
beyond the SM is to assume that $Y_d$ and $Y_u$ are the only 
sources of flavor symmetry  breaking also in the NP model.
To implement and interpret this hypothesis in a consistent way, 
we can assume that ${\mathcal G}_{q}$ is a good symmetry and 
promote $Y_{u,d}$ to be non-dynamical fields ({\em spurions}) with 
non-trivial transformation properties under  ${\mathcal G}_{q}$:
\begin{equation}
Y_u \sim (3, \bar 3,1)~,\qquad
Y_d \sim (3, 1, \bar 3)~.\qquad
\end{equation}
If the breaking of the symmetry occurs at very high energy scales, 
at low-energies we would only be sensitive to the background values of 
the $Y$, i.e.~to the ordinary SM Yukawa couplings. 
The role of the Yukawa in breaking the flavor symmetry becomes 
similar to the role of the Higgs in the the breaking of the 
gauge symmetry. However, in the case of the Yukawa we don't 
know (and we do not attempt to construct) a dynamical 
model which give rise to this symmetry breaking.

Within the effective-theory approach 
to physics beyond the SM introduced above, 
we can say that an effective theory satisfies the criterion of
Minimal flavor Violation in the quark sector 
if all higher-dimensional operators,
constructed from SM and $Y$ fields, are invariant under CP and (formally)
under the flavor group ${\mathcal G}_{q}$.

According to this criterion one should in principle 
consider operators with arbitrary powers of the (dimensionless) 
Yukawa fields. However, a strong simplification arises by the 
observation that all the eigenvalues of the Yukawa matrices 
are small, but for the top one, and that the off-diagonal 
elements of the CKM matrix are very suppressed. 
Working in the basis where 
$Y_{d}$ is diagonal, where $Y_d=\lambda_d~, \qquad  Y_u=V^\dagger\lambda_u$,
we have 
\be
\left[  Y_u (Y_u)^\dagger \right]^n_{i\not = j} ~\approx~ 
y_t^n V^*_{it} V_{tj}~.
\ee
As a consequence, in the limit where we neglect light quark masses,
the leading $\Delta F=2$ and $\Delta F=1$ 
flavor Changing Neutral Current (FCNC) 
amplitudes get exactly 
the same CKM suppression as in the SM: 
\begin{eqnarray}
  \mathcal{A}(d^i \to d^j)_{\rm MFV} &=&   (V^*_{ti} V_{tj})^{\phantom{a}} 
 \mathcal{A}^{(\Delta F=1)}_{\rm SM}
\left[ 1 + a_1 \frac{ 16 \pi^2 M^2_W }{ \Lambda^2 } \right]~,
\\
  \mathcal{A}(M_{ij}-\bar M_{ij})_{\rm MFV}  &=&  (V^*_{ti} V_{tj})^2  
 \mathcal{A}^{(\Delta F=2)}_{\rm SM}
\left[ 1 + a_2 \frac{ 16 \pi^2 M^2_W }{ \Lambda^2 } \right]~.
\label{eq:FC}
\end{eqnarray}
where the $\mathcal{A}^{(i)}_{\rm SM}$ are the SM loop amplitudes 
and the $a_i$ are $\mathcal{O}(1)$ real parameters. The  $a_i$
depend on the specific operator considered but are flavor 
independent. This implies the same relative correction 
in $s\to d$, $b\to d$, and  $b\to s$ transitions 
of the same type: a key prediction which can be tested 
in experiment.

As pointed out in Ref.~\cite{Buras:2000dm}, within the MFV
framework several of the constraints used to determine the CKM matrix
(and in particular the unitarity triangle) are not affected by NP.
In this framework, NP effects are negligible not only in tree-level
processes but also in a few clean observables sensitive to loop
effects, such as the time-dependent CPV asymmetry in $B_d \to \psi
K_{L,S}$. Indeed the structure of the basic flavor-changing coupling
in Eq.~(\ref{eq:FC}) implies that the weak CPV phase of $B_d$--$\bar
B_d$ mixing is arg[$(V_{td}V_{tb}^*)^2$], exactly as in the SM.  
This construction provides a natural (a posteriori) justification 
of why no NP effects have been observed in
the quark sector: by construction, most of the clean observables
measured at $B$ factories are insensitive to NP effects in the MFV
framework.  

Given the built-in CKM suppression, the bounds on 
higher-dimen\-sio\-nal operators in the MFV framework
turns out to be in the TeV range. 
This can easily be understood by the discussion 
in Sect.~\ref{sect:DF2bounds}: the MFV bounds 
on operators contributing to $\epsilon_K$ and $\Delta m_{B_d}$ 
are obtained from  
Eq.~(\ref{eq:boundsDF2}) setting $|c_{ij}| =  |y_t^2 V_{3i}^* V_{3j}|^2$.
From this perspective we could say that the MFV 
hypothesis provides a solution 
to the NP flavor problem.

\subsubsection*{General considerations on MFV and beyond}

\begin{enumerate}
\item[I.]
The idea that the CKM matrix rules the strength of  FCNC transitions also beyond the SM 
is a concept that requires a proper justification. The CKM matrix 
represents only one part of the problem: a key role in
determining the structure of FCNCs is also played  by quark masses, or 
better by the Yukawa eigenvalues. 
In this respect, the MFV criterion provides the maximal protection 
of FCNCs (or the minimal violation of the flavor symmetry), since the full 
structure of Yukawa matrices is preserved. Moreover, 
 the MFV criterion  is based on a renormalization-group-invariant 
 symmetry argument: we cannot have a more minimal breaking  of 
 the flavor symmetry since the Yukawa couplings are already present in the 
 SM and, via quantum corrections, necessarily propagate their 
 symmetry-breaking structure to the whole tower of effective operators.
 
\item[II.]
Although MFV seems to be a natural solution to the NP flavor problem, we are still far from having proved the validity of this hypothesis from data. 
A proof of the MFV hypothesis can be achieved only with a positive evidence of NP exhibiting the flavor-universality pattern
predicted by MFV (same relative correction in $s\to d$, $b\to d$, and $b\to s$ transitions of the same type). 
This could happens, for instance, via precise measurements of the rare decays 
 $B_s\to\mu^+\mu^-$ and $B_d\to\mu^+\mu^-$. 
Conversely, an evidence of NP in 
flavor-changing transitions not respecting the MFV pattern (e.g.~an evidence of $\cB(B_d\to\mu^+\mu^-)$
 well above its SM prediction) would not only imply the existence of physics beyond the SM, 
but also the existence of new sources of flavor symmetry breaking beyond the Yukawa couplings.

\item[III.]
If we assume the hierarchical pattern of the SM Yukawa couplings has a dynamical origin, the MFV ansatz cannot be an 
exact property of the theory valid to all energy scales: the MFV anastz simply ``postpone'' the problem of generating 
the Yukawa couplings to some unspecified high-scale dynamics. 
In this respective, an interesting alternative are flavor symmetries (and corresponding symmetry-breaking ansatz) 
acting only on the light families~\cite{Barbieri:1995uv,Pomarol:1995xc,Kagan:2009bn,Barbieri:2011ci,Barbieri:2012uh,Blankenburg:2012nx,Isidori:2012ts,Glioti:2024hye}. 
The largest of such groups is $U(2)^5$, which is defined in full analogy to $U(3)^5$ in Eq.~(\ref{eq:Gtot}),
but considering only  the first two generations of fermions~\cite{Barbieri:2011ci,Barbieri:2012uh,Blankenburg:2012nx,Isidori:2012ts}.
This symmetry provides a ``natural'' explanation of why 
third-generation Yukawa couplings are large (being allowed by the symmetry). Moreover, contrary to the MFV case, 
it allows us to build an EFT  where all the breaking terms are small, offering a more precise power counting 
for the effective operators~\cite{Faroughy:2020ina,Greljo:2022cah}.
With a suitable choice of the (small) symmetry breaking terms, the hypothesis of an approximate $U(2)^5$ 
symmetry can be as effective as MFV in protecting FCNCs~\cite{Barbieri:2011ci,Isidori:2012ts}.
 
\item[IV.]
The usefulness of the MFV ansatz is closely linked to the theoretical expectation of NP in the TeV range.
This expectation follows from a  ``natural'' stabilization of the Higgs sector. In the last few years this expectation
is in growing tension with the lack of direct signals of NP at the LHC. As we will discuss next, going beyond MFV this tension 
can be minimised.

\end{enumerate}

\subsection{Flavor non-universal gauge interactions}

The MFV hypothesis can view as an attempt  two separate the flavor problem and the electroweak hierarchy problem.
The basic idea is to stabilise the Higgs sector (hence addressing the electroweak hierarchy problem)
via some sort of flavor-universal new physics (such as supersymmetry or Higgs-compositeness)
close to the electroweak scale, while postponing the origin of flavor dynamics to higher scales. 
This scale separation   is possible because the flavor hierarchies are associated to marginal
operators (the Yukawa interaction), which do not indicate a well-defined energy scale. However, the hypothesis of 
flavor-universal dynamics just above the electroweak scale has become less and less natural in the last few years:
the absence of direct signals of new physics at the LHC has pushed well above 1~TeV the bounds on new degrees of freedom with $O(1)$  flavor-universal couplings to the SM fields. This fact unavoidably worsen the stabilisation of the electroweak scale, independently of the possible solution to the flavor problem.

An important point to notice is that the stringent bounds from direct searches are not derived from direct couplings of the new physics to the Higgs or the top quark, but rather by its couplings to the light SM fermions, which play a minor role in the stability of the Higgs mass. As illustrated in the previous sections, a similar conclusion applies to the indirect constraints on new physics couplings derived from precision low-energy measurements: are the effective operators involving light SM fermions to be strongly constrained, not those involving only third-generation fields 
 (see~\cite{Allwicher:2023shc} for a detailed quantitative analysis). These general observations seems to indicate that 
flavor and electroweak problems are interrelated and should not be addressed separately, but rather in combination.

This reasoning is at the base of the growing interest on UV completions of the SM with flavor non-universal gauge interactions. 
The main idea is addressing the origin of the flavor hierarchies via gauge interactions which are manifestly 
non universal in flavor, or better via the {\em flavor deconstruction} in the UV~\cite{Arkani-Hamed:2001nha,Craig:2011yk}
of the flavor-universal gauge structure of the SM. 
An interesting prototype for this class of UV completions is the ${\rm PS}^3$ model
proposed in Ref.~\cite{Bordone:2017bld}, namely the flavor deconstruction of the semi-simple  gauge group 
 \begin{equation}
 {\rm PS}=\SU(4)\times \SU(2)_L\times \SU(2)_R\,,
 \end{equation}
 proposed by Pati and Salam in 1974~\cite{Pati:1974yy}. 
After this attempt, various closely related proposals and a few interesting alternatives 
have been discussed in the literature~\cite{Greljo:2018tuh,Fuentes-Martin:2020pww,Fuentes-Martin:2020bnh,Fuentes-Martin:2022xnb,FernandezNavarro:2022gst,FernandezNavarro:2023rhv,Davighi:2022fer,Davighi:2022bqf,Davighi:2023iks,Davighi:2023evx,Barbieri:2023qpf,Davighi:2023xqn,Covone:2024elw}.  
Initially,  these constructions were motivated by  the hints of violations of flavor universality in $B$ decays mentioned in the previous lecture.
However,  their interest goes well  beyond this phenomenological aspect: they provide a natural explanation for the observed hierarchical pattern of the SM Yukawa coupling and for the 
appearance of the global $\UU(2)^5$ flavor symmetry as accidental symmetry of the 
extended gauge sector around the TeV scale~\cite{Davighi:2023iks,Barbieri:2023qpf}.

The basic idea of flavor deconstruction
 works as follows~\cite{Davighi:2023iks}. Consider the following symmetries acting on all the SM 
fermions:\footnote{Here hypercharge is decomposed into a chiral component, $T^3_R$, and a vector-like component, $B-L$ (under which the Higgs is neutral), via the relation $Y= T^3_R +(B-L)/2$. }
 $\SU(2)_L$, $\UU(1)_{B-L}$, and $\UU(1)_{R}\equiv U(1)_{T^3_R}$.
Denoting $G$ any of this groups, or the product of some of them,  we assume that above some energy threshold, $\Lambda_3 > v$,
the theory is invariant under the local symmetry 
\be
G^{[12]} \times G^{[3]}\,.
\label{eq:FlavDec}
\ee
Here $G^{[i]}$ indicates that the symmetry group $G$ acts only on the generation $i$. 
At energy scales below $\Lambda_3$ this UV symmetry undergoes the spontaneous symmetry breaking 
\be
G^{[12]} \times G^{[3]}  \to G^{[123]} \equiv  G^{[{\rm univ}]} \supset G_{\rm SM}\,,
\label{eq:FlavDec2}
\ee
such that flavor universality naturally emerges at low energies. 
Actually the symmetry breaking process in Eq.~(\ref{eq:FlavDec2}) is the last step in the deconstruction process, which presumably occurs around the TeV scale. At higher scales also $G^{[12]}$ 
could be deconstructed into groups acting differently on first and second generation. 
%At the TeV scale, the presence of a difference gauge symmetry for the first two generations and the third one leads to the appearance of a global $\UU(2)^n$ flavor symmetry (the index $n$ denote the number of fermions charged under $G$). 

The deconstruction  of   $\SU(2)_L$, $\UU(1)_{B-L}$, or $\UU(1)_{R}$, forbid certain Yukawa couplings at the renormalisable level.
More precisely, considering each of the groups separately, leads to the following allowed pattern for the Yukawa couplings:

\begin{table}[h]
\begin{center}
\begin{tabular}{cccc}
& 	{\footnotesize $\SU(2)_L^\light \times \SU(2)_L^\heavy$}  
& 	{\footnotesize $\UU(1)_{B-L}^\light \times \UU(1)_{B-L}^\heavy$}
&	{\footnotesize $\UU(1)_R^\light \times \UU(1)_R^\heavy$} \\
&	{\footnotesize [with $H \sim {\bf (1, 2)}$]}
& &	{\footnotesize [with $H\sim (0,-1/2)$]}  \\[2mm]
$Y_f \sim$ 
&	$\qquad\begin{pmatrix} 0&0&0\\ 0&0&0\\ \times&\times&\times \end{pmatrix}\qquad$
&	$\qquad\begin{pmatrix} \times&\times&0\\ \times&\times&0\\ 0&0&\times \end{pmatrix}\qquad$
&	$\qquad\begin{pmatrix} 0&0&\times\\ 0&0&\times\\ 0&0&\times \end{pmatrix}\qquad$ 
\end{tabular}
\end{center}
\end{table}
\vskip - 0.3 cm
\noindent
As can be seen, deconstructing any pair of these groups implies that only the $(Y_f)_{33}$ entries of the Yukawa couplings 
are allowed, hence  a global $U(2)^5$ emerges as accidental symmetry of the gauge sector. 
The various options in the different models presented in the literature differ 
essentially on which of these groups are chosen, on the final embedding in the UV, and on
the symmetry breaking pattern. Choosing $\UU(1)_{B-L}$ and $\UU(1)_{R}$, 
merged into a family-dependendent hypercharge, corresponds to the 
minimal option~\cite{FernandezNavarro:2023rhv,Davighi:2023evx,Barbieri:2023qpf}.

\begin{figure}[t]
\centering
\includegraphics[width=0.87\textwidth]{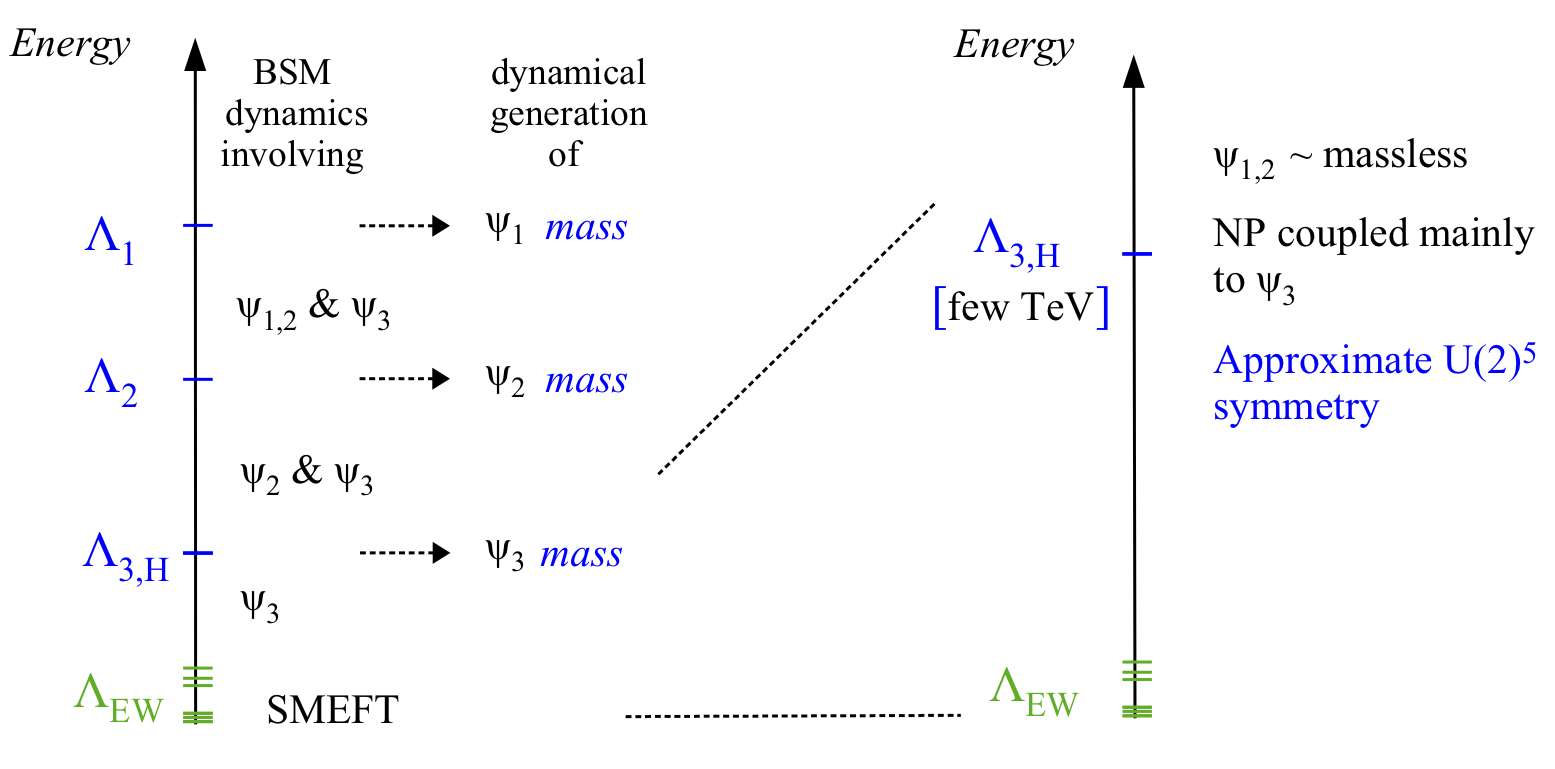} 
\caption{\label{fig:Multiscale} Schematic representation of the possible multi-scale construction at the origin of the SM flavor hierarchies.  }
\end{figure}

This approach also connects to the older, and more general, idea of a UV completion of the  SM based on a (flavored) multi-scale construction~\cite{Dvali:2000ha,Panico:2016ull,Allwicher:2020esa,Barbieri:2021wrc}, 
where the masses for the light generations are generated at increasingly higher scales (see Fig.~\ref{fig:Multiscale}).
In this context the naturalness of the Higgs mass can be preserved, or, more accurately, the tuning is minimized, if the first layer of new physics enters near the TeV scale, with subsequent layers separated by a few orders of magnitude~\cite{Allwicher:2020esa}. This of course does not rule out the possibility that supersymmetry or compositeness 
play a role in screening the scalar sector from dynamics at even heavier scales; however, if utilised, these general stabilisation mechanisms  could manifest at higher scales given the low-scale stabilisation provided by the lighter flavor non-universal layer of new-physics.

\subsection{Present data and future prospects}

As shown in~\cite{Davighi:2023iks}, if the flavor-deconstruction idea presented above is implemented in models with a 
semi-simple embedding in the UV, addressing also the hypercharge quantisation problem, the number of possible options is 
quite limited. Most important, all viable models share common features.  A notable common prediction is the expectation 
of a TeV-scale vector leptoquark ($U_1$ field), coupled mainly to the third generation, resulting from a 
unification \`a la Pati and Salam for the third family.

\begin{figure}[p]
\centering
\includegraphics[scale=0.65]{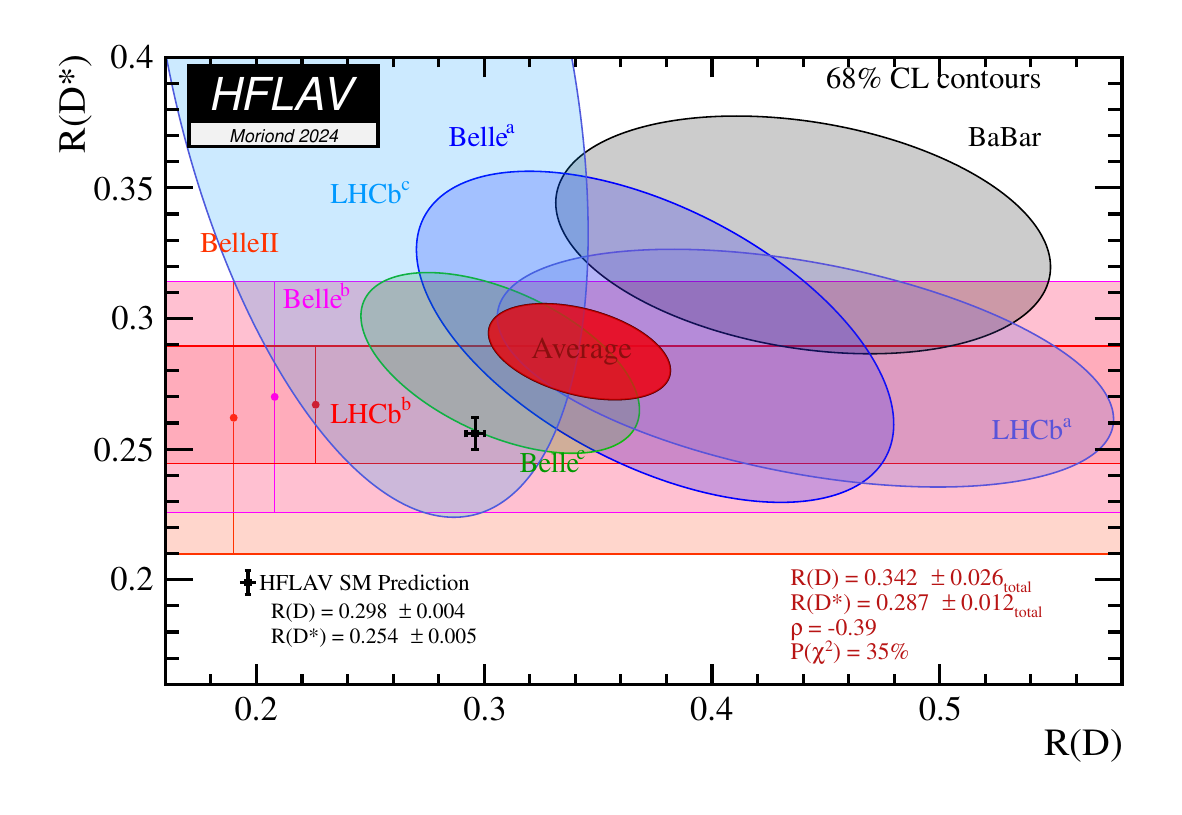}
\caption{Status of the experimental results on the LFU ratios $R_D$ and $R_{D^*}$  (colored bands and ellipses) vs.~SM predictions (black cross) in winter 2024~\cite{HeavyFlavorAveragingGroupHFLAV:2024ctg}. The red ellipse indicate the $68\%$~C.L. region resulting from the global average of the experimental results. }
\label{fig:RD}
\vskip 1.0 cm 
    \includegraphics[width=0.45\linewidth]{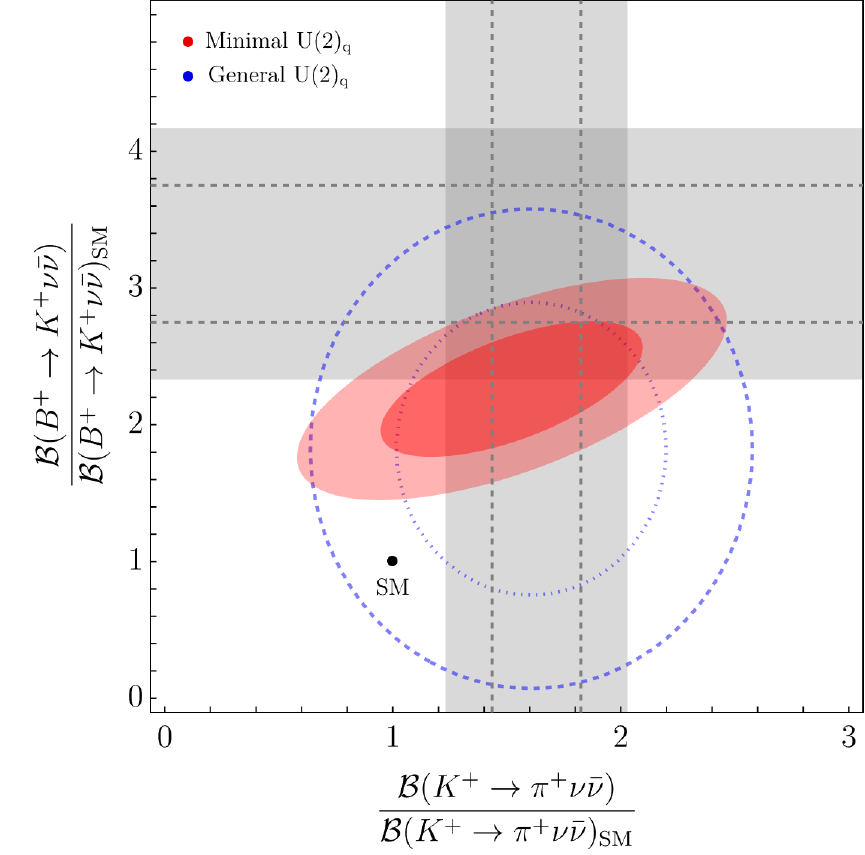}~
    \includegraphics[width=0.47\linewidth]{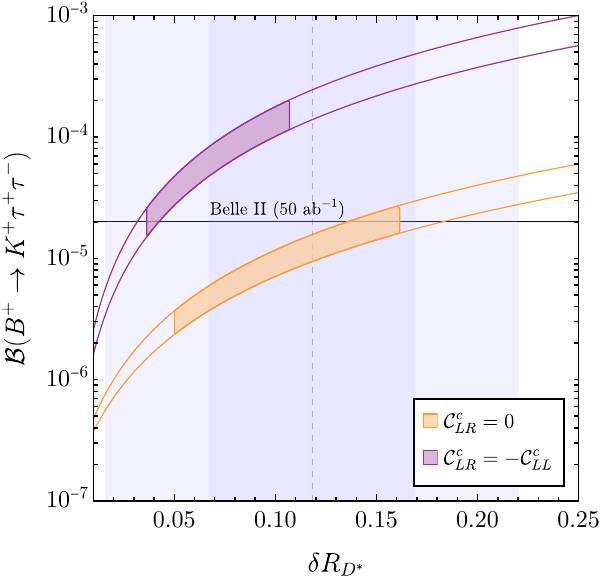}
    \caption{Examples of low-energy observables which can exhibit significant deviations from the SM in models with TeV scale NP
    coupled  mainly to the third generation, addressing the $R_{D^{(*)}}$ anomaly. Left: $\cB(B^+\to K^+\bar \nu \nu)$~vs.~$\cB(K\to \pi^+\bar \nu \nu)$~\cite{Allwicher:2024ncl}.
    The red areas denote the parameter regions favored at $1\sigma$ and $2\sigma$ from a global fit in the limit of minimal $U(2)^5$ breaking; the gray bands indicate current experimental constraints; gray lines highlight near-future projections. Right: $\cB(B\to K^+ \tau^+ \tau^-)$ vs.~the relative shift in $R_{D^*}$ assuming  Pati-Salam unification at the TeV scale~\cite{Aebischer:2022oqe}.
    The blue bands denote the present 1 and 2$\sigma$ value of $R_{D^*}$; orange (violet) bands correspond to a pure left-handed (left-right symmetric) interaction of the $U_1$
     to third-generation fermions. }
    \label{fig:RareBvaria} 
\end{figure}

A $U_1$ field coupled mainly to the third generation, respecting an approximate $U(2)^5$ flavor symmetry, can explain
the persisting tension between data and SM predictions in the 
LFU ratios $R_D$ and $R_{D^*}$ probing $\tau/\mu$ universality in $b\to c\ell\nu$ transitions~\cite{Alonso:2015sja,Calibbi:2015kma,Barbieri:2015yvd}.
According to HFLAV~\cite{HeavyFlavorAveragingGroupHFLAV:2024ctg} 
(see also~\cite{Bordone:2024weh,Martinelli:2023fwm}),  the latest global fit indicate an enhancement 
over the SM predictions of the $\tau$ rates (over light leptons) of about 10\%, with a significance of around 3$\sigma$ (see Figure~\ref{fig:RD}).
The SM contribution to these processes appears already at the tree level, and is suppressed only by $|V_{cb}|$. 
Hence a NP interpretation of this anomaly necessarily points toward a sizeable amplitude, corresponding to an
effective scale of at most a few TeV. 

As recently discussed in~\cite{Allwicher:2024ncl}, TeV scale NP respecting an approximate $U(2)^5$ flavor symmetry and affecting mainly semilptonic processes, is supported by two additional and independent sets of observables. First, by the enhancement with respect to the SM of $B^+\to K^+\nu\bar\nu$~\cite{Belle-II:2023esi} and $K^+ \to \pi^+ \nu\bar\nu$~\cite{NA62:2024pjp} rates, which are particularly interesting since they involve third-generation neutrinos in the final state. Second, by the  deficit observed in the (lepton-universal) value of $C_9$ determined from exclusive $b\to s\mu\bar \mu$ transitions. As discussed in the second lecture, this is subject to non-negligible theoretical uncertainties; 
still, present data indicate a tension with the SM of at least $2\sigma$ (see~\cite{Alguero:2022wkd,Bordone:2024hui,Isidori:2024lng}). None of these effects are particularly convincing on their own. However, when considered together, they form a coherent and interesting picture.

If the $U_1$ hypothesis is correct, more particles and more signals should be accessible in the short term at both low and high energies
First of all, the $U_1$ leptoquark cannot be alone~\cite{Baker:2019sli}. The minimal consistent gauge group hosting this massive vector 
is $\SU(4)^{[3]}\times \SU(3)^{[12]} \times \SU(2)_L \times \UU(1)$~\cite{DiLuzio:2017vat}. Its breaking down to the SM 
implies the presence of three sets of massive vectors:  in addition to 
the $U_1$,  also a color octet $G$  (denoted {\em coloron}, behaving like a heavy gluon) and a $Z^\prime$ (singlet under the SM) are necessarily present. All these fields are coupled dominantly to third-generation fermions, with similar mass ($M_V$) 
and couplings ($g_4$).
As mentioned above, if we require the $U_1$ to address the $R_{D^{(*)}}$, then 
$M_V/g_4 \in [1,2]$~TeV.  Such range is very close to present bound 
from direct searches. In particular, the most promising channels to detect these massive vectors are  
$pp\to \bar\tau\tau$ ($t$--channel  $U_1$ and $s$--channel $Z^\prime$ exchange~\cite{Faroughy:2016osc,Baker:2019sli}) and $pp\to \bar t t$ ($s$--channel $G$ exchange~\cite{Baker:2019sli,Cornella:2021sby}).

A general expectation of these new states and, more generally, of any model with flavor deconstruction at the TeV scale, 
are deviations from the SM predictions in various electroweak precision observables  (typically at the few per-mil level). 
As shown in~\cite{Davighi:2023evx,Davighi:2023xqn,Stefanek:2024kds} by means of explicit examples,  FCC-ee would allow to extensively probe the natural parameter space of such  class of models via electroweak observables.
Last but not least, a necessary ingredient of this class of models are vector-like fermions (i.e.~fermions where left- and right-handed
components have the same gauge quantum numbers, allowing a Dirac-mass term). These fermions are responsible for the 
heavy-light mixing of the chiral fermions (or the structure of the CKM matrix), as it happens in a wider  class of motivated SM extensions~\cite{Botella:2016ibj}. The interplay of massive vectors and vector-like fermions  give rise to a series of additional
low-energy signatures which are within the reach of the present generation of flavor-phyics experiments (see Fig.~\ref{fig:RareBvaria}).
 These include lepton-flavor violating decays such as  $B\to K^{(*)}\bar \tau \mu$,  
$\tau \to \phi \mu$,  or $\tau \to 3\mu$ close to present bounds (see e.g.~Ref.~\cite{Cornella:2021sby});
large enhancements over the SM predictions for rare $B$ decays into $\tau^+\tau^-$ 
 pairs~\cite{Capdevila:2017iqn}; $O(10-30\%)$ deviations over the SM predictions for $\cB(B\to K^{(*)}\bar \nu \nu)$~\cite{Fuentes-Martin:2020hvc}  and    $\cB(K^+ \to \pi^+\bar \nu \nu)$~\cite{Crosas:2022quq}.

  \section{Conclusions}
  
Flavor physics presents two intriguing open questions. First, the origin of the peculiar structures observed in quark and lepton mass matrices. Second, the absence of significant deviations from the SM in numerous precise low-energy flavor physics measurements.

In these lectures, I have explored these questions and examined their implications for constraining physics beyond the SM.  Although it is not clear whether these two issues are related, as I discussed in the final part of the lectures, an interesting possibility is that they are.  By accepting this connection, we can conceive of a new physics scenario that addresses both the origin of the flavor hierarchies and the electroweak hierarchy problem  while satisfying all experimental constraints.  The key ingredient is the assumption of a multiscale completion of the SM, with a first threshold around the TeV scale.  The new physics present at this threshold, which is not universal in flavor, is coupled mainly to the third generation and is responsible for the stabilization of the Higgs sector.  Notably, in this framework is possible to explain some of the hints of deviations from the SM currently observed in various semileptonic $B$-meson  decays. The individual significance of these deviations is not very high, but when considered together they form a rather interesting picture. 

On the other hand, it is fair to state that the paradigm of new physics at the TeV scale faces significant challenges, given the lack of deviation from the SM in high-energy searches. Even with the perspective that NP lies beyond the reach of direct searches, flavor physics remains a powerful tool to probe it. 
As shown with a few explicit examples, flavor-changing processes have the potential to probe NP at much higher scales than those directly accessible in present and near-future high-energy experiments.

As I have pointed out throughout these lectures, one of the most fascinating aspects of flavor physics today is the great increase in experimental data in many processes expected in the coming years. In the next two decades the increase in precision in flavor physics, and the corresponding progress in testing the SM via flavor-changing processes, is potentially greater than in other areas of high-energy physics. This makes flavor physics a very exciting field, with the potential for surprising discoveries in the near future.

\section*{Acknowledgements}
I am grateful to the organizers of the High Energy Physics School 2024 Asia-Europe-Pacific for the invitation to this interesting school, for the warm hospitality and for the wonderful organization. This project has received funding from the Swiss National Science Foundation (SNF) under contract 200020 204428.

\addcontentsline{toc}{section}{\numberline {6}References}

\footnotesize{
\bibliographystyle{JHEP}
\bibliography{refs}
}

\end{document}